\documentclass[12pt,preprint]{aastex}

\begin{document}

%\slugcomment{Submitted to ApJ}

%
%MACROS
%
\newcommand{\vect}[1]{\ensuremath{\mbox{\boldmath $#1$}}}
\def\Ae {A_\bme}
\def\bby  {{\bf y}}
\def\beq{\begin{equation}}
\def\beqa{\begin{eqnarray}}
\def\beqan{\begin{eqnarray*}}
\def\bby  {{\bf y}}
\def\bm {{\mathop{\rm Mag}\nolimits }}
\def\bmo  {{\bf 0}}
\def\br  {{\bf r}}
\def\bs  {{\bf s}}
\def\bt  {\vect{\theta}}
\def\btcl  {  {\vect{\theta}}_{\rm cl}   }
\def\bphicl { {\vect{\varphi}}_{\rm cl} }
\def\btclc  {  {\vect{\theta}}_{\rm loc}   }
\def\btclcfs  {  {\vect{\theta}}_{\rm loc}^{\rm fs}   }
\def\btclcout  {  {\vect{\theta}}_{\rm loc}^{\rm out}   }
\def\btclcin {  {\vect{\theta}}_{\rm loc}^{\rm in}   }
\def\btclcoutfs  {  {\vect{\theta}}_{\rm loc}^{\rm fs, out}   }
\def\btclcinfs  {  {\vect{\theta}}_{\rm loc}^{\rm fs, in}   }
\def\btclcperpfs {  {\vect{\theta}}_{\rm loc}^{\rm fs, \perp}}
\def\btclcoutld  {  {\vect{\theta}}_{\rm loc}^{\rm ld, out}   }
\def\btclcinld  {  {\vect{\theta}}_{\rm loc}^{\rm ld, in}   }
\def\btclfs  {  {\vect{\theta}}_{\rm cl}^{\rm fs}   }
\def\btclus  {  {\vect{\theta}}_{\rm cl}^{\rm us}   }
\def\btclcusout  {  {\vect{\theta}}_{\rm loc}^{\rm us, out}   }
\def\btclcusin {  {\vect{\theta}}_{\rm loc}^{\rm us, in}   }
\def\btclh  {  {\hat{\vect{\theta}}}_{\rm cl}   }
\def\btin  {  {\vect{\theta}}^{\rm in}   }
\def\btout  {  {\vect{\theta}}^{\rm out}   }
\def\bton  {  {\vect{\theta}}^{\rm on}   }
\def\bTcl {\delta\vect{\theta}_{\rm cl}}
\def\bTclc {\delta\vect{\theta}_{\rm loc}}
\def\bu  {\vect{u}}
\def\bucn  {{\vect{u}}_{\rm cn}}
\def\cD { {\cal{D}} }
\def\cP { {\cal{P}} }
\def\cQ { {\cal{Q}} }
\def\cR { {\cal{R}} }
\def\cT { {\cal{T}} }
\def\det {\mathop{\rm det}\nolimits}
\def\dls{D_{ls}}
\def\dol{D_{ol}}
\def\dos{D_{os}}
\def\dsum  {\displaystyle\mathop{\sum}}
\def\dtj{\Delta \theta_{jump}}
\def\eeq{\end{equation}}
\def\eeqa{\end{eqnarray}}
\def\eeqan{\end{eqnarray*}}
\def\eq#1{equation~(\ref{#1})}
\def\Eq#1{Eq.[\ref{#1}]}
\def\event{MACHO 97-BLG-28}
\def\grad {\mathop{\rm grad}\nolimits}
\def\jac  {\mathop{\rm Jac}\nolimits}
\def\hbu{\hat{\bu}}
\def\hbt{\hat{\bt}}
\def\hmu  {\hat{\mu}}
\def\hs {\Delta u^*_{1,\vdash}}
\def\hsfd {\Delta u^{\rm fold,d}_{1,\vdash}}
\def\hsfs {\Delta u^{\rm fold,s}_{1,\vdash}}
\def\hsinell {\Delta u^{\rm in, \ell}_{1,\vdash}}
\def\hsina {\Delta u^{\rm in, 1}_{1,\vdash}}
\def\hsinb {\Delta u^{\rm in, 2}_{1,\vdash}}
\def\hsinc {\Delta u^{\rm in, 3}_{1,\vdash}}
\def\hsout {\Delta u^{\rm out}_{1,\vdash}}
\def\mfs{{\mu^{\rm fs}}}
\def\muas{\mu{\rm as}}
\def\mul{\mu_{\rm loc}}
\def\mulfs{\mu_{\rm loc}^{\rm fs}}
\def\mulldout{\mu_{\rm loc}^{\rm ld, out}}
\def\mulldin{\mu_{\rm loc}^{\rm ld, in}}
\def\mulus{\mu_{\rm loc}^{\rm us}}
\def\mulusout{\mu_{\rm loc}^{\rm us, out}}
\def\mulusin{\mu_{\rm loc}^{\rm us, in}}
\def\mulout{\mu_{\rm loc}^{\rm out}}
\def\muloutfs{\mu_{\rm loc}^{\rm fs, out}}
\def\mulinfs{\mu_{\rm loc}^{\rm fs, in}}
\def\muloutperp{\mu_{\rm loc}^{\perp}}
\def\muloutperpfs{\mu_{\rm loc}^{\rm fs, \perp}}
\def\muloutperpus{\mu_{\rm loc}^{\rm us, \perp}}
\def\muloutperpld{\mu_{\rm loc}^{\rm ld, \perp}}
\def\mulons{\mu_{\rm loc}^{\rm fold, s}}
\def\mulond{\mu_{\rm loc}^{\rm fold, d}}
\def\mulin{\mu_{\rm loc}^{\rm in}}
\def\mut{\mu_{\rm tot}}
\def\mtfs{{\mu_{\rm tot}^{\rm fs}}}
\def\mtld{{\mu_{\rm tot}^{\rm ld}}}
\def\mtus{{\mu_{\rm tot}^{\rm us}}}
\def\pa  {\partial}
\def\rd { {\rm d} }
\def\rK { {\rm K} }
\def\ru  {{\mbox{\rm u}}}
\def\rv  {{\mbox{\rm v}}}
\def\Re { R_{\rm E} }
\def\RR  {{\bf R}}
\def\sign {\mathop{\rm sign}\nolimits}
\def\Sn { {S_N}}
\def\tcc{t_{c}}
\def\te{t_{\rm E}}
\def\th {\theta}
\def\thetae{\theta_{\rm E}}
\def\thetas{\theta_\epsilon}
\def\ttp  {{\rm p}}
\def\ttq  {{\rm q}}
\def\vs {\Delta u_{2,\perp}}
\def\vst {\Delta \theta_{2,\perp}}
\def\ui {\vect{\hat{\rm \i}}}
\newcommand{\bme}{ {\mbox{\boldmath $\eta$}} }
\newcommand{\bma}{ {\mbox{\boldmath $\alpha$}} }

\title{Gravitational Microlensing Near Caustics II: Cusps}

\author{B. Scott Gaudi\altaffilmark{1,2} and A. O. Petters\altaffilmark{3,4}}

\altaffiltext{1}{School of Natural Sciences, 
Institute for Advanced Study, Princeton, NJ 08540, gaudi@sns.ias.edu}
\altaffiltext{2}{Hubble Fellow}
\altaffiltext{3}{Department of Mathematics, Duke University,
Durham, NC 27708, petters@math.duke.edu}
\altaffiltext{4}{Bass Society of Fellows, Duke University}

\begin{abstract}
We present a rigorous, detailed study of the generic, quantitative
properties of gravitational lensing near cusp catastrophes.
Concentrating on the case when the individual images are unresolved,
we derive explicit formulas for the total magnification and centroid
of the images created for sources outside, on, and inside the cusped
caustic.  We obtain new results on how the image magnifications scale with
respect to separation from the cusped caustic for arbitrary source
positions.  Along the axis of symmetry of the cusp, the total
magnification $\mu$ scales as $\mu \propto u^{-1}$, where $u$ is the
distance of the source from the cusp, whereas perpendicular to this
axis, $\mu \propto u^{-2/3}$.  When the source passes through a
point $\bu_0$ on a fold
arc abutting the cusp, the image centroid has a jump discontinuity; we present a formula
for the size of the jump in terms of the local derivatives of the lens
potential and show that the magnitude of the jump scales
as $|u^0_1|^{1/2}$ for $|u^0_1| \ll 1$, where
$|u^0_1|$ is the horizontal distance between $\bu_0$ and
the cusp.  The total magnifications for a small extended
source located both on, and perpendicular to, the axis of symmetry are
also derived, for both uniform and limb darkened surface brightness
profiles.  We find that the difference in magnification
between a finite and point source is $\la 5\%$ for separations of
$\ga 2.5$ source radii from the cusp point, while the effect
of limb-darkening is $\la 1\%$ in the same range. Our predictions for the
astrometric and photometric behavior of both pointlike and finite
sources passing near a cusp are illustrated and verified using
numerical simulations of the cusp-crossing Galactic binary-lens event
MACHO-1997-BUL-28.  Our results can be applied to any microlensing
system with cusp caustics, including Galactic binary lenses and
quasar microlensing; we discuss several possible applications of our
results to these topics.
\end{abstract}


\section{Introduction}

Over the past twenty years, gravitational lensing has grown from
a mere curiosity to an important component of a large and diverse
set of fields in astronomy.  Its ubiquity is due at least in part to 
the fact that its effects are observable over a wide range of scales.  
This has enabled astronomers to use lensing to study everything from
the smallest compact objects, to the largest structures in the universe,
and almost everything in between. 
Despite the diversity of applications of gravitational lensing, the
mathematical description of the phenomenon itself is both relatively
tractable, and universal.  In almost all cases, gravitational lensing can
be described by a two-dimensional mapping from a lens plane to light source 
plane.  Once this mapping is specified, all of the properties of
a gravitational lens can be derived in principle.  The {\it observable}
properties of lensing, however, depend on the phenomenon to which
it is applied.  Therefore, lensing is
traditionally divided into an number of different regimes, which are
delimited by the observables.  For example, the term
microlensing is typically applied to the case when multiple images
occur, but are not resolved.  When multiple images are created
by a gravitational lens, the separation between these images is typically
of order the Einstein ring radius,
\begin{equation}
\thetae=\sqrt{{{4 G M}\over c^2 D} },
\label{eqn:thetae}
\end{equation}
where $M$ is the mass of the lens, $D\equiv
\dos\dol/\dls$, and $\dos$, $\dol$, and $\dls$ are the distances
from observer to source, observer to lens, and lens to source,
respectively.  Thus, the term microlensing is applied when $\thetae$ is less than the resolution.  In this case, all one can measure is the collective
behavior of all the images created by the lens, i.e.,\ the total magnification,
and the position of the center-of-light (centroid) of the image.  In fact, if the observer, source, and lens were not in relative motion, than the 
individual image magnifications and positions would be fixed, and even these
properties would not be measurable.  The relative positions of the observer, lens, and source, and
thus the magnification and centroid, are expected
to vary on time scales of order the Einstein ring crossing time,
\begin{equation}
\te = {\thetae \dol\over v_\perp},
\label{eqn:te}
\end{equation}
where $v_\perp$ is the transverse speed of the lens relative to the
observer-source line-of-sight.  Fortunately, in the two regimes where
microlensing has been discussed, the typical values of $\thetae$,
$\dol$, and $v_\perp$ result in reasonable time scales.  Typical
values for the lens mass $M$, relative source-lens distance $D$, 
transverse velocity $v_\perp$, and the resulting typical values 
for $\thetae$ and $\te$ for both Local Group and cosmological microlensing
are given in Table 1.  Also shown are typical values
for the radius  $\theta_*$ of the source's emission region, this radius
in units of $\thetae$, $\rho_*\equiv \theta_*/\thetae$, and
the time it takes the lens to cross the source, $t_*\equiv \rho_* \te$.  
These latter parameters will be relevant to the discussion of finite
source effects in \S\ref{sec:fsources}.  For the Local Group, a typical Einstein radius
crossing time is $\te={\cal O}(100~{\rm days})$, whereas for cosmological 
microlensing, $\te={\cal O}(10~{\rm years})$.  

\begin{table}[t]
\begin{center}
\begin{tabular}{c|cccccccc}
\tableline
\multicolumn{1}{c}{ } &
\multicolumn{1}{c}{$M$} &
\multicolumn{1}{c}{$D$} &
\multicolumn{1}{c}{$\thetae$} &
\multicolumn{1}{c}{$v_\perp$} &
\multicolumn{1}{c}{$\te$} &
\multicolumn{1}{c}{$\theta_*$} &
\multicolumn{1}{c}{$\rho_*$} &
\multicolumn{1}{c}{$t_*$} \\
\tableline
Local Group  & $1 M_\odot$ & $10~{\rm kpc}$ & $1{\rm mas}$ & $100~{\rm km~s^{-1}}$ & $100~{\rm days}$ & $1\muas$ & $10^{-3}$ & $0.1~{\rm days}$\\
\tableline
Cosmological  & $1 M_\odot$ & $1~{\rm Gpc}$ & $3\muas$ & $500~{\rm km~s^{-1}}$ & $10~{\rm years}$ & $0.1\muas$ & $0.03$ & $100~{\rm days}$\\
\tableline
\end{tabular}
\end{center}
\tablenum{1} {\bf Table 1} Typical Microlensing Parameters. \\
\label{tbl:table1}
\end{table}

Of exceptional importance in microlensing is the existence of
caustics:  positions in the light source plane corresponding
to the critical values of the lens mapping.
%positions in the light source plane where the mapping become
%critical, and the determinant of the Jacobian of the lens mapping
%vanishes.  
On caustics, at least one image is formed that has formally
infinite magnification (for a point-source).  When a source
crosses a caustic, both the total magnification and centroid of all
the images exhibit instantaneous, discontinuous jumps.  These jumps
are averaged out over the finite source size; however, it is
generically true that large gradients in the magnification and
centroid exist near caustics.  Furthermore, microlensing caustics have several
important and useful properties.  First, the large magnification
results in a large photon flux from the source.  Second, the large
gradient in the magnification and centroid with respect to source
position effectively implies high angular resolution.  Finally, the
highly localized nature of the high-magnification and large
centroid-shift regions created by caustics results in characteristic,
and easily-recognizable features, in both astrometric and photometric
microlensing curves.  Many authors have suggested exploiting these
properties of caustics to study a number of astrophysical
applications, i.e.,\ stellar multiplicity \citep{mandp1991}, stellar
atmospheres \citep{gould2001}, individual microlens mass
measurements \citep{gandg2002}, 
microlens mass functions \citep{wwt2000a}, properties of the
emission regions of quasars
\citep{wps1990,ak1999,fw1999,wwt2000b}, and lens transverse
velocities \citep{wwt1999}.  See Gaudi \&
Petters (2002; hereafter Paper I) for a more thorough discussion 
of the uses of astrometric and photometric microlensing observations
in the presence of caustics in both Local Group and cosmological contexts.

Although the caustic curves of gravitational lenses 
exhibit an enormously rich and diverse range
of properties, it can be shown rigorously that each stable  
lensing map has only 
two types of caustic singularities: folds and cusps
(Petters, Levine, \& Wambsganss
2001, p. 294).  Each of these two types of singularities have generic 
and universal properties and, in particular, each can be described
by a polynomial mapping from the lens plane to light source plane. 
The coefficients of these mappings depend on local derivatives
of the dimensionless surface potential of the lens.  In Paper I, we 
used the mapping for a fold singularity to derive the observable 
properties of gravitational lensing near folds, paying particular
attention to the case of microlensing, in which the images are unresolved.
We derived analytic expressions for the total magnification and centroid
shift near a generic, parabolic, fold caustic.  We then showed how 
these expressions reduce to those for the more familiar linear fold, which
lenses a nearby source into two equal magnification, opposite parity images
whose total magnification is proportional to $u^{-1/2}$, where $u$ is the 
distance of the source to the fold caustic.  We then generalized these
results to finite source sizes.  Finally, we compared our analytic results 
to numerical simulations of the Galactic
binary-lens event OGLE-1999-BUL-23, in which the source was observed to cross a fold
caustic.  We found excellent qualitative agreement between our analytic and semi-analytic
expressions for the photometric and astrometric behavior near a fold
caustic, and our detailed numerical simulations of the second fold caustic
crossing of OGLE-1998-BUL-23.

In this paper, we present a similarly detailed study of the generic,
quantitative properties of microlensing near cusps.  Although fold
caustic crossings are expected and observed to dominate the sample of
caustic crossings in Galactic binary events \citep{gandg1999,ggh2002,
alcock2000}, cusp crossings will nevertheless represent a
non-negligible fraction of all caustic crossing events.  In fact, at
least two cusp crossing events have already been observed, the
Galactic bulge events \event ~\citep{albrow1999a}, and
MACHO-1997-BLG-41 \citep{alcock2000, albrow2000}.  It is interesting
to note that the analyses and modeling of these events were performed
using entirely numerical methods.  As we discuss in some detail (see \S\ref{sec:apply}),
we believe that the analytic results derived here are particularly
amenable to the analysis of \event~ and similar events.  In the
cosmological context, the role of cusps versus folds is less clear,
due primarily to the more complicated structure of the caustics
themselves.  However, it has been shown that, in the limit of high
magnifications, the inclusion of cusps alters the form of the
total microimage magnification cross section, due to the lobe of
high-magnification close to and outside the cusped caustic
\citep{sw92}.  Lensing near cusps has been substantially less
well-studied than lensing near folds, and as a result, 
useful analytic expressions for the observable properties are few.  Previous
studies have focused almost exclusively on the magnification of the images
created by a cusp singularity \citep{sw92, mao92, zak95, zak99}.  Here
we study all the observable properties of gravitational microlensing
near cusps, including the photometric (total magnification) and astrometric
(light centroid) behavior, for both point sources and extended sources with arbitrary
surface brightness profiles.

The layout of this paper is as follows.  In \S\ref{sec:analytic}, we
derive analytic expressions for the image positions, magnification,
and light centroid for sources near a generic cusp.  In
\S\ref{sec:general} we define the observable microlensing properties.
In \S\ref{sec:local}, we start with the generic
expression for the mapping near a cusp, and derive all the properties
for the local images for sources exterior to (\S\ref{sec:exterior}), on
(\S\ref{sec:on}), and interior to (\S\ref{sec:interior}) the caustic.
We generalize the discussion to include images not associated with the
cusp in \S\ref{sec:global}, and study extended sources in
\S\ref{sec:fsources}.  In \S\ref{sec:mb9728}, we illustrate the
observable behavior near a cusp by numerically simulating the
Galactic binary-lens cusp-crossing event \event, and directly compare
these numerical results with our analytic expressions in
\S\ref{sec:compare}.  In \S\ref{sec:apply}, we suggest
several possible applications in both local group and cosmological 
contexts.  We summarize and conclude in \S\ref{sec:summary}. We note that, for the sake of completeness, we
include some results that have been presented elsewhere.  Combined
with the results from Paper I, the results presented here describe the
observable properties of gravitational microlensing near all stable
singularities.

\section{Analytical Treatment\label{sec:analytic}}

\subsection{General Equations for Magnification and Image Centroid\label{sec:general}}

The lens equation due to a gravitational lens with 
potential $\psi$
is given in dimensionless form as follows:
\beq
\label{eq-le}
\bu = \bt - \bma (\bt).
\eeq
Here $\bt = \br/(\thetae \dol)$ and
$\bu = \bs/({\theta_{\rm E} \dos})$,
where $\vect{r}$ and $\vect{s}$ denote the proper vector
positions in the lens and light source planes, respectively, and
$\bma (\bt) \equiv {\vect{\nabla}} \psi (\bt).$
Note that if $\kappa$ is the surface mass density of
the lens  in units of the critical density
$\Sigma_{cr}\equiv (c^2 \dos)/(4 \pi G \dol \dls)$,
then  
the gravitational lens potential $\psi$ is related
to $\kappa$ by
\beq
\psi (\bt)  
= \frac{1}{\pi} \int_{\rm R^2}d{\bt'} \kappa(\bt')\ln |\bt-\bt'|.
\label{eqn:potential}
\eeq
The lens equation induces a lensing map $\bme$
from the lens plane $L$ into the light source plane $S$ 
defined by
$\bme (\bt) = \bt - \bma (\bt)$.
The critical curves of $\bme$ are defined to be
the locus of all critical points of
$\bme$, i.e., the set of solutions $\bt_c$ of the
equation
\beq
\det[\Ae](\bt_c) =0,
\label{eqn:deta}
\eeq
where $\Ae$ is the Jacobian matrix of $\bme$.
The caustics of $\bme$ are its critical values, i.e.,
the set of all points $\bu_c = \bme (\bt_c)$.

For a fixed source position $\bu$, the solutions
in $L$ of the lens equation 
determine the
lensed images of $\bu$.  
Suppose that a  source at $\bu$ has 
$N(\bu)$ images.  The  
magnification of the $i$th image
$\bt_i = \bt_i (\bu)$ is given by
\beq
\mu (\bt_i) = \frac{1}{|\det[\Ae](\bt_i)|}. 
\label{eqn:magimagei}
\eeq
Then the total magnification is 
\beq
\label{def-mtot}
\mut (\bu) =
\dsum_{i=1}^{N(\bu)}  \mu (\bt_i).
\eeq
For a source at $\bu$, the {\it image centroid} 
or {\it center-of-light}, denoted by 
$\btcl (\bu)$, and {\it shifted image centroid}, which
is denoted by $\delta \btcl$, 
are defined respectively as follows: 
\beq
\label{def-colc}
\btcl (\bu) = \frac{1}{\mut (\bu)} \dsum_{i=1}^{N(\bu)} \mu (\bt_i) \bt_i,
\hspace{.5in}
\delta \btcl (\bu) = \btcl (\bu) - \bu.
\eeq
The image centroid is defined analogously to a center
of mass.  However, the image centroid is much more complicated since
images can suddenly be created or annihilated as the source
moves over caustics. For the case of gravitational microlensing,
the total magnification and shifted centroid are observables, with
the latter expected to be 
accessible to, e.g.\ the {\it Space Interferometry Mission} (SIM).

\subsection{Local Case: Point Sources Near a Cusp\label{sec:local}}

A rigorous analytical study of the total magnification (photometry)
and image centroids (astrometry) for most physically reasonable lens
models (e.g., binary point mass lenses) is very difficult.  
Our goal here, as in Paper I, is to study the local, generic behavior of 
these observables near an {\it arbitrary, but fixed,} critical point of a given type.
In Paper I, we studied the behavior near a fold singularity; here
we study the behavior near a cusp.  In order to study
the local behavior near singularities, we Taylor expand
the scalar potential $\psi$ in the neighborhood of the singularity, 
keeping terms to a specified order in the image position $\bt$.  We note
that the order at which one chooses to truncate the expansion is somewhat arbitrary;
higher accuracy can always be achieved by keeping terms of higher order (at the expense of higher complexity, of course).  However, it is generically true that 
fold critical points are characterized by partial derivatives of the lensing map
up to second order, while cusps are classified by partials up to third order (Petters et al.\ 2001, pp. 370-371).  As a result, the Taylor expansion of the potential is usually
taken up to second order for the fold and third order for the cusp.
See Schneider et~al.\
(1992, pp. 187-188) and particularly Petters et~al.\ (2001, pp. 341-353) 
for more discussion on truncation of
Taylor expansions of the potential.  After introducing 
an orthogonal change in coordinates, one can thus obtain a simpler equation
that approximates the quantitative, generic behavior of the total
magnification and image centroid of sources near a critical point. Consider a
generic lensing map $\bme$ whose caustics have cusps. 
In such a case, there are at least two cusps since the total
number of cusps is always {\it even} 
(see Petters 1995, p. 4282, for a cusp-counting formula).
In this section we will fix one of the cusps and study the behavior of $\bme$ in the
vicinity of that cusp.  

\subsubsection{Local Case: Magnification and Centroid of 
Images Associated with the Cusp}
\label{subsec-local-case}

Without loss of generality, we can assume that the coordinates $\bt$
on the lens plane $L$ and $\bu$ on the light source plane $S$ are
translated so that the cusp lies at the origin $\bmo$ of $S$ and $\bme
(\bmo) = \bmo$.  There is an orthogonal change of the coordinates
$\bt$ and $\bu$ which is the same in the lens and light source planes
such that the lensing map $\bme$ can be approximated by the
following simpler mapping in a neighborhood of the origin (Petters et~al.\
2001, pp. 341-353; Schneider et~al.\ 1992, p. 193):
\beq
\label{eq-le-cusp}
u_1 = c \th_1 + \frac{b}{2} \th^2_2, \qquad 
u_2 = b \th_1 \th_2 + a \th_2^3,
\eeq
where
\beq
a = -\frac{1}{6} \psi_{2222}(\bmo), 
\quad b = -\psi_{122} (\bmo)
\neq 0, \quad c = 1 - \psi_{11} (\bmo) \neq 0,
\quad  2ac - b^2 \neq 0.
\label{eqn:coeff}
\eeq
Note that in (\ref{eq-le-cusp}) we are abusing notation somewhat by
using the notation  
$\bt = (\th_1, \th_2)$ and  $\bu = (u_1, u_2)$
for the global lens equation  (\ref{eq-le}) 
to express the local lens equation.
For the explicit relationship between 
the local coordinates in equation (\ref{eq-le-cusp}) and the 
global ones in equation (\ref{eq-le}), see
pp. 344-345 of Petters et al. (2001).
It is also important to add that the partials in
(\ref{eqn:coeff}) are with respect
to the original global coordinates of the lens equation.
Figure \ref{fig:example1} illustrates the coordinate systems and 
basic properties of lensing near a cusp.  For this figure, we
have adopted the coefficients $a=1.69, b=2.45$, and $c=2.0$.  These 
are the appropriate values for the cusp in the observed event {\event}.  See \S\ref{sec:compare}.

We shall now use the lens mapping determined by equation (\ref{eq-le-cusp}) 
to study the approximate
behavior of the magnification and  image centroid near a cusped caustic curve.
The magnification matrix $A$ for the cusp is
the Jacobian matrix of \eq{eq-le-cusp}, 
\beq
\label{eq-Jac-matrix}
A = \left(\begin{array}{lc}
           c        &  b  \th_2 \\
           b  \th_2 &  b  \th_1 + 3a  \th_2^2 
           \end{array}
    \right),
\eeq
which has the determinant
\beq
\label{eq-detA}
\det A = bc \th_1 + (3ac - b^2) \th_2^2.
\eeq
The cusp critical point of the mapping in \eq{eq-le-cusp}
is at the origin of the lens plane and
mapped to a cusp caustic point at the origin of the light source plane.
By \eq{eq-detA}, the critical curve is a parabola,
\beq
\label{eq-cc}
\th_1 = \frac{b^2  - 3ac}{bc} \th_2^2.
\eeq
Equations (\ref{eq-le-cusp}) and (\ref{eq-cc}) yield 
that points $\bu_c = (u_{c,1}, u_{c,2})$ on the caustic
are given by
\beq
\label{eq-caustic-para}
u_{c,1} = \frac{3(b^2-2ac)}{2b} \th_2^2,
\qquad 
u_{c,2} = \frac{b^2-2ac}{c} \th_2^3,
\eeq 
where  $\th_2$ acts as a parameter along the caustic.
The magnitude of the curvature of the caustic is
$$
|k| = \frac{|u_{c,1}^\prime \ u_{c,2}^{\prime \prime} - 
      u_{c,2}^\prime \ u_{c,1}^{\prime \prime}|}{
        [ (u_{c,1}^\prime)^2 + (u_{c,2}^\prime)^2]^{3/2}}
= \frac{|b\ c|}{3(b^2 + c^2)\ |b^2 - 2 ac|} \ \frac{1}{|\th_2|},
$$
where the primes indicate differentiation with respect to
$\th_2$.   The curvature diverges as the cusp is 
approached ($\th_2 \rightarrow 0$).

By \eq{eq-caustic-para}, the caustic can also be
described as the level curve
\beq
\label{eq-C}
C (\bu) \equiv  \frac{8b^3}{27c^2(2ac - b^2)} u_1^3 + u_2^2 =0.
\eeq
The origin is a {\it positive cusp} if
$2ac > b^2 $ and a {\it negative cusp} if $2ac < b^2$.  
Without loss of generality, {\it we shall assume that
the cusp is positive.}  For this case, both 
$a$ and $c$ have the same sign.  The conditions $C(\bu) > 0$,
$C(\bu) = 0$, and $C(\bu) < 0$ determine
points $\bu$ outside, on, and inside the cusped caustic curve,
respectively.

Let us determine the local lensed images and their magnifications
for sources near the cusped caustic. 
The expression for $u_1$ in \eq{eq-le-cusp} yields
\beq
\label{eq-x1}
\th_1 = \frac{u_1}{c}  - \frac{b}{2c} \th_2^2.
\eeq
Plugging \eq{eq-x1} into the expression for $u_2$ in \eq{eq-le-cusp}, we
obtain a cubic equation for $\th_2$ in terms of the source position
$(u_1, u_2)$:
\beq
\label{eq-cubicpol}
f(\th_2) \equiv \th_2^3 + \ttp \ \th_2 + \ttq  =  0,
\eeq
where
\beq
\ttp = \ttp (u_1)= \frac{2 b }{2ac - b^2} u_1,
\qquad \ttq = \ttq (u_2) = -\frac{2c}{2ac - b^2} u_2.
\label{eqn:defpq}
\eeq
In other words, the lensed images are points $\bt = (\th_1,\th_2)$ with
$\th_1$ given by \eq{eq-x1} and $\th_2$  a real root of
\eq{eq-cubicpol}.  The magnification of
each lensed image $\bt$ of $\bu$ is then
\beq
\mu (\bt) = \frac{1}{|\det A|},
\label{eqn:muva}
\eeq
where
\beq
\label{eq-detA-y}
\det A = b u_1  + \frac{3}{2} (2ac - b^2) \th_2^2
\eeq
with 
$\th_2$ a real zero of $f$.

For a source at angular vector position $\bu$, 
let $\bt_* (\bu) = (\th^*_1 (\bu), \th^*_2 (\bu))$
be any lensed image of $\bu$ associated with the cusp,
i.e.,
\beq
\bt_* (\bu) = \left(\frac{u_1}{c}  - \frac{b}{2c} [\th^*_2 (\bu)]^2, 
\th^*_2 (\bu) \right),
\label{eqn:thetas}
\eeq
where $\th^*_2 (\bu)$ is a solution of cubic equation (\ref{eq-cubicpol}). 
Then \eq{eq-caustic-para} yields that 
\beq
u^*_{c,1} = \frac{3(b^2-2ac)}{2b} (\th^*_2)^2,
\qquad
u^*_{c,2} = \frac{b^2-2ac}{c} (\th^*_2)^3,
\label{eqn:ucdefs}
\eeq
is a point on the caustic.     
It is important to note that
even for a source position $\bu = \bu_c$ at a caustic point,
there is no guarantee that the caustic point
$\bu^*_c = (u_{c,1}^*, u_{c,2}^*)$  equals $\bu_c$
(see \Eq{eq-hsfs-n} below).
Now, consider the tangent line to the cusp, which happens 
to coincide with the $u_1$-axis (due to our coordinate choice).
We shall also refer to this tangent line as the axis of the cusp, 
since it coincides with the axis of symmetry of the cusp.
Note that by (\ref{eq-Jac-matrix}) the magnification
matrix $A(\bmo)$ at the origin,  maps  the entire lens plane into
the previous tangent line.
Define
the {\it horizontal distance} 
between $\bu = (u_1,u_2)$ and $\bu^*_c = (u^*_{c,1}, u^*_{c,2})$ 
to be the distance between $\bu$ and $\bu^*_c$ along the direction
parallel to the tangent line to the cusp.
Denoting the horizontal distance by
$\hs$, we obtain
\beq
\hs  \equiv | u_1 - u^*_{c,1}| = | u_1 - \frac{3(b^2-2ac)}{2b} (\th^*_2)^2|.
\eeq
By \eq{eq-detA-y}, we see that 
{\it the magnification of $\bt_*$ is  inversely proportional to
the horizontal distance between $\bu$ and $\bu^*_c$:}
\beq
\label{eq-mag-hd} 
\mu (\bt_*) =  \frac{1}{|b| \hs}.
\eeq
We shall see that if the source
is on the  $u_1$-axis, then the magnification of
the image $\bt_*$ (only one image occurs locally for sources outside
the caustic) is proportional to $u^{-1}$, where
$u$ is the distance between the source and 
cusp (see Eqs.[\ref{eq-mag-out-u1-axis}] and
[\ref{eq-magin-u1}] below). 
If the source
is on the  $u_2$-axis, then the magnification of
$\bt_*$  (again, only one image occurs locally)
is proportional to $u^{-2/3}$
(see Eqs. \ref{eq-mag-out-u2-axis}).
In both cases, the horizontal
distance coincides with the ordinary distance.

We note that in Paper I, we showed that the
magnification of an image $\bt_*$ associated with
the fold caustic is inversely proportional to the
square root of the {\it perpendicular distance} 
of the source from the fold.
For a generic local fold lensing map that
maps a fold critical point at the origin
to a fold caustic point at the origin,
the perpendicular distance is defined relative to
the direction perpendicular to the tangent
line to the fold caustic at the origin.
Note that the magnification matrix $A (\bmo)$
for the fold also maps the entire lens plane into
the tangent line.

Now, the reals zeros of the polynomial $f$ in \eq{eq-cubicpol} are characterized
using the  
discriminant of $f$, i.e.,
\beq
\label{eq-discriminant}
D (\bu) \equiv \left(\frac{\ttp}{3}\right)^3 +  \left(\frac{\ttq}{2}\right)^2
= \frac{c^2}{(2ac-b^2)^2} C(\bu).
\eeq
By \eq{eq-discriminant}, the conditions $D(\bu) > 0$,
$D(\bu) = 0$, and $D(\bu) < 0$ also determine, respectively,
points $\bu$ outside, on, and inside the cusped caustic curve.
It will be seen that for a source outside, on, or inside the
cusped caustic,  there are locally one, two, or three images, respectively.
These images lie in the neighborhood of the origin, which is a point
on the critical curve.
The magnification and centroid of the images associated with the cusp
will be denoted by $\mul$ and
$\btclc$, respectively.,  and called the {\it local magnification} and
{\it local image centroid}.  We shall now determine 
$\mul$ and $\btclc$ for source positions near the cusp that
are outside, on, and inside
the caustic --- see equations (\ref{eq-loc-mag}) and
(\ref{eq-loc-centroid}) for a summary.

\subsubsection{Local Case: Source Outside the Cusp\label{sec:exterior}}

For a source position $\bu$ outside
and near the cusp (i.e., $D(\bu) > 0$), equation
(\ref{eq-cubicpol}) has one real
root,
namely,
\beqa
\th^{\rm out}_2 (\bu)
&=&
\left( -\frac{\ttq}{2} + \sqrt{D(\bu)}\right)^{\frac{1}{3}}
+
\left( -\frac{\ttq}{2} - \sqrt{D(\bu)}\right)^{\frac{1}{3}},
\label{eq-x2out-1}
\\
&=& \left( \frac{c}{2ac - b^2} \right)^{\frac{1}{3}}
\left[
\left( u_2 + \sqrt{C(\bu)}\right)^{\frac{1}{3}}
+
\left( u_2 -  \sqrt{C(\bu)}\right)^{\frac{1}{3}}
\right].
\label{eq-x2out}
\eeqa
Consequently, there is one lensed image  (locally)
\beq
\label{eq-out-li}
\bt^{\rm out} (\bu) =
\left(\frac{u_1}{c} - \frac{b}{2c} [\th^{\rm out}_2 (\bu) ]^2,
\  \th^{\rm out}_2 (\bu) \right).
\eeq
To determine the parity of $\btout(\bu)$, it suffices
to find the sign of $\det A (\btout (\bu))$ for a single source position
outside the caustic.  A source lying on
the $u_2$-axis either above or
below the origin is outside the caustic curve
(since $C (0,u_2) = u_2^2 >0$ for $u_2 \neq 0$). Furthermore,
$ \th^{\rm out}_2 (0,u_2) = (-\ttq)^{\frac{1}{3}},$
which is nonzero for $u_2 \neq 0$.  It follows that
\beq
\det A [\btout (0,u_2)] =
\frac{3}{2} (2ac - b^2)  (-\ttq)^{\frac{2}{3}} >0,
\eeq
where we used the fact that the cusp is positive (i.e., $2ac - b^2
>0$).
Hence, the image $\btout$ has positive parity.
In the case of lensing by $n$-point masses, there are no maximum
images (Petters 1992) and so $\bt^{\rm out}$ is a minimum image.

\begin{figure}
\plotone{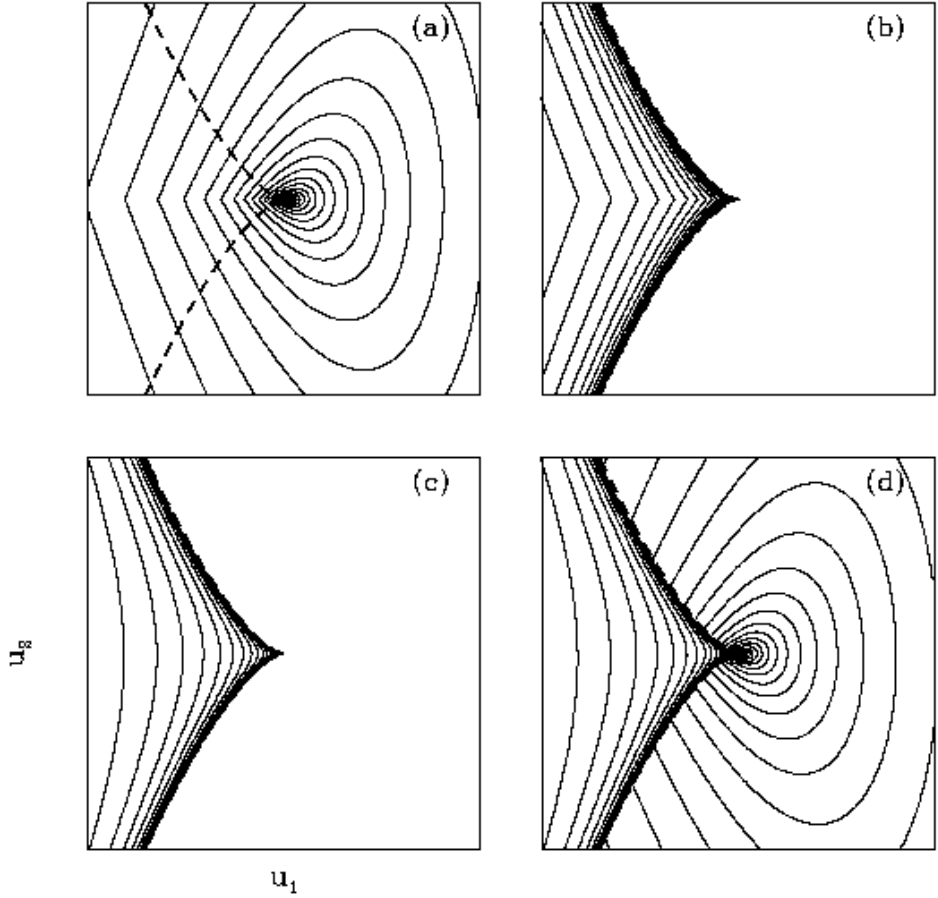}
\caption{ 
Contour plots of the magnification of the various images as a function
of source position, $\bu$.  In each panel, contours are in the
interval $\log(\mu)=-1,1$, in logarithmic steps of 0.1, and the dashed
line is the caustic.  (a) Contours of magnification of the
non-divergent, positive-parity image, as a function of $\bu$.  Outside
the caustic, this corresponds to $\mulout(\bu)$.  Inside the caustic,
this corresponds to $\mu^{\rm in}_1 (\bu)$ for $u_2>0$, and $\mu^{\rm
in}_2 (\bu)$ for $u_2<0$.  (b) Contours of the divergent,
positive-parity image, which corresponds to $\mu^{\rm in}_1 (\bu)$ for
$u_2<0$, and $\mu^{\rm in}_2 (\bu)$ for $u_2>0$. (c) Contours of the
divergent, negative-parity image, which corresponds to $\mu^{\rm in}_3
(\bu)$.  (d) Contours of the total magnification.  See text for the
definitions of $\mulout$, $\mu^{\rm in}_1$, $\mu^{\rm in}_2$, and
$\mu^{\rm in}_3$
}
\label{fig:conmat}
\end{figure}

The magnification of the image $\btout (\bu)$ is
\beq
\label{eq-mag-out}
\mulout (\bu) \equiv \mu (\bt^{\rm out}(\bu)) =  \frac{1}{|b| \hsout},
\eeq
where
\beq
\label{eq-hs-out}
\hsout \equiv  |u_1  -  \frac{3}{2b} (b^2 - 2ac) [\th^{\rm out}_2 (\bu)]^2|.
\eeq
As
$\bu$ approaches the cusp at $\bmo$ from outside the caustic
(i.e., the constraint $D (\bu)>0$ holds
for $\bu \neq \bmo$),
we have  $\th^{\rm out}_2 (\bu) \rightarrow 0$.  Consequently,
the magnification $\mulout$ continuously increases without bound as
the cusp is approached from outside. This is unlike the case
for a source crosses a fold caustic since then a sudden, infinite,
discontinuous-jump occurs in the magnification (for  point sources).
The continuous increase in $\mulout$ is due to a lobe of high-magnification
outside the cusp (see Fig.\ \ref{fig:conmat}, and also pp.\ 334-335, 480-485 of Petters et~al.\ 2001).

We now consider the magnification of sources on the axes, exterior to the cusp.  
For sources on the $u_2$-axis and outside the cusp,  
we have $\ttp=0$ and (thus) $C(\bu)=u_2^2$.   The resulting
image position is therefore 
\beq
\label{eq-out-u2-axis}
\bt^{\rm out} (0,u_2) =
\left[-\frac{b}{2c} \left(\frac{2c}{2ac-b^2}u_2\right)^{2/3}, \left(\frac{2c}{2ac-b^2}u_2\right)^{1/3}\right],
\eeq
and the magnification is given by,
\beq
\label{eq-mag-out-u2-axis}
\mulout (0,u_2) \equiv \mu (\bt^{\rm out}(0,u_2)) = \left(\frac{u_{r2}}{u_2}\right)^{2/3},
\eeq
where we have defined the characteristic rise scale $u_{r2}$ for trajectories
along the $u_2$-axis,
\beq
\label{eq-uc2def}
u_{r2}\equiv\left[\frac{27}{2}c^2\left(2ac-b^2\right)\right]^{-1/2}.
\eeq
Thus, the magnification diverges as $u_2^{-2/3}$ for $u_2 \rightarrow 0$.

If the source is on the $u_1$-axis and outside the cusp
(i.e., $u_1 >0$), then $\ttq=0$.  
Equation
(\ref{eq-x2out-1}) yields $\th^{\rm out}_2 (u_1,0) =0$, so
the image position is simply,  
\beq
\label{eq-out-u1-axis}
\bt^{\rm out} (u_1,0) =
\left(\frac{u_1}{c}, 0\right),
\eeq
and the magnification is given by,
\beq
\label{eq-mag-out-u1-axis}
\mulout (0,u_2) \equiv \mu (\bt^{\rm out}(u_1,0)) = \frac{|u_{r1}|}{u_1},
\eeq
where we have defined the rise scale $u_{r1}$,
\beq
\label{eq-uc1def}
u_{r1}\equiv b^{-1}.
\eeq
Hence, the magnification diverges as $u_1^{-1}$  for $u_1 \rightarrow 0$.

\begin{figure}
\plotone{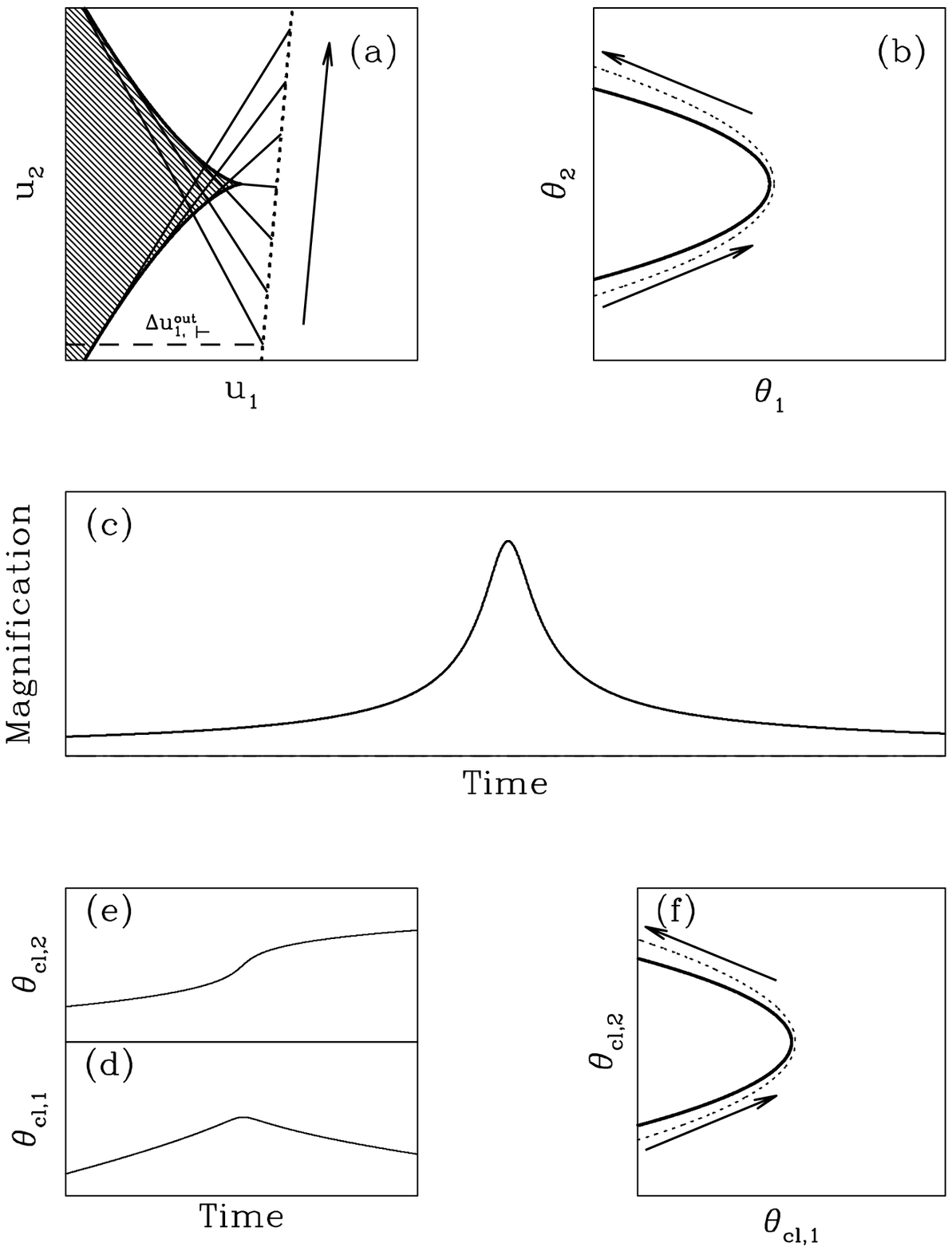}
\caption{ 
Illustration of the basic properties of photometric and astrometric
microlensing near a cusp, for a source trajectory which passes near,
but exterior to, the cusp point.  (a) The heavy solid line is the
caustic, the shaded area indicates the region interior to the caustic,
where three images are formed; one image occurs for sources
outside the bounded shaded region.  The dotted line is the source
trajectory.  Each sloped line connects a source position $\bu$
to the point $\bu_{c}^*$ on the caustic.
The $u_1$-difference
between $\bu_{c}^*$ and $\bu$ is the
horizontal distance $\hsout\equiv |u_{1}-\bu_{c,1}^*|$.  
The magnification of an image is $\propto (\hsout)^{-1}$.  See text.
(b) The solid line is the parabolic critical curve,
whereas the dotted line denotes the image's trajectory.  (c) The
magnification of the image is shown in (b) as a function of time.  The
solid line is the total magnification.  (d) The
$\theta_{cl,1}$-component of the centroid of all the images as a
function of time.  (e) The $\theta_{cl,2}$-component of the centroid
as a function of time.  (f) The solid line is the critical curve,
whereas the dotted line shows the path of the centroid of all the
images.  
}
\label{fig:example1}
\end{figure}

For a source moving along a path
$\bu (t)$ (not necessarily rectilinear)
lying outside the
cusp, the local image centroid follows the motion of the
lensed image $\bt^{\rm out} (\bu (t))$ (see the path of
$\btout$ in Figure~\ref{fig:example1}),
i.e.,
\beq
\label{eq-colout}
\btclc^{\rm out} (\bu (t))
= \bt^{\rm out} (\bu(t)).
\eeq
Observe that $\btclc^{\rm out} (\bmo) = \bmo$.
Figures~\ref{fig:example1} (b) and (f) illustrate
the trajectories of the image and centroid, respectively,
for a source in rectilinear motion. 

Let us now consider the behavior of the lensed image
$\btout$ when the source at $\bu$ approaches the caustic.
First, as $\bu$ approaches the cusp $\bmo$ from
outside, we obtain $\btout (\bu) \rightarrow \bmo$.
Second, at a point $\bu_0 = (u_1^0, u_2^0)$ on one of the fold arcs meeting
the cusp, \eq{eq-x2out-1} shows that 
$\th^{\rm out}_2 (\bu_0) = 2 (-\ttq_0/2)^{1/3}.$
Since $D(\bu_0) =0$, it follows from
\eq{eq-discriminant} that
\beq
\label{eq-thout2}
\frac{-4\ttp_0}{3} = 4 \left(\frac{\ttq_0}{2}\right)^{\frac{2}{3}}
= \left[ \th^{\rm out}_2 (\bu_0) \right]^2,
\eeq
where $\ttp_0 = \ttp(u_1^0)$ and
 $\ttq_0 = \ttq(u_2^0)$. Note
that \eq{eq-thout2} implies
$-\ttp_0 >0$, which yields
\beq
\label{eq-bu01}
b u_1^0 < 0
\eeq
(since $2ac- b^2 >0$).
Hence, the limiting position of $\bt^{\rm out} (\bu)$ as 
$\bu  \rightarrow \bu_0$ from outside the caustic is 
\beq
\label{eq-li-fold-out}
\btout (\bu_0) =
\left(   \frac{u^0_1}{c} + \frac{2b}{3c}\ \ttp_0,
               \  2   \left(\frac{- \ttq_0}{2}\right)^{\frac{1}{3}}\right).
\eeq

\subsubsection{Local Case: Source on the Caustic\label{sec:on}}

Assume that the source is on the caustic (i.e., $D(\bu) = 0$).
If the source is at the cusp point $\bu = \bmo$, then
\beq
\ttp = \ttq =0,
\eeq
and \eq{eq-cubicpol} has a real triple root, $\bt_2 (\bmo)
=0$.  This yields an infinitely magnified lensed image at the origin
of the lens plane.  Consequently, for a source on the cusp, the local
image centroid is at the origin (as can be seen from
\Eq{eq-colout}):
\beq
\label{eq-colcusp-a}
\bmo = \bt^{\rm out} (\bmo) = \btclc^{\rm on} (\bmo).
\eeq

If the source is at a point $\bu_0 = (u_1^0, u^0_2)$ on
one of the fold arcs abutting the cusp (i.e.,
$D (\bu_0) =0$ and $\bu_0 \neq \bmo$), then
\beq
\ttp_0 \ttq_0 \neq 0,
\eeq
so \eq{eq-cubicpol} has two real roots, one a simple zero and
the other a double zero:
\beq
\label{eq-th2fsd}
\th^{\rm fold, s}_2 (\bu_0 )
= 2 \left(\frac{- \ttq_0}{2}\right)^{\frac{1}{3}},
\qquad
\th^{\rm fold, d}_2 (\bu_0)
= - \left(\frac{-\ttq_0}{2}\right)^{\frac{1}{3}}.
\eeq
The corresponding lensed images are as follows
(since $D (\bu_0) =0$):
\beqa
\label{eq-li-foldd}
\bt^{\rm fold, d} (\bu_0)
&\equiv&
\left(\frac{u_1^0}{c} + \frac{b}{6c}\ \ttp_0,  
              -   \left(\frac{- \ttq_0}{2}\right)^{\frac{1}{3}}  \right), \\
\label{eq-li-folds}
\bt^{\rm fold, s} (\bu_0)
&\equiv&  
\left(   \frac{u_1^0}{c} + \frac{2b}{3c}\ \ttp_0,
                2   \left(\frac{- \ttq_0}{2}\right)^{\frac{1}{3}}\right)
= \bt^{\rm out} (\bu_0).
\eeqa
Note that $\bt^{\rm fold, d} (\bmo) = \bmo = \bt^{\rm fold, s} (\bmo)$
and the rightmost equality in \eq{eq-li-folds} 
is a consequence of \eq{eq-li-fold-out}.

Using \eq{eq-detA-y}, \eq{eq-th2fsd}, and $D (\bu_0) =0$, we get
\beqa
\label{eq-detAs}
\det A (\bt^{\rm fold, s} (\bu_0))
& = &
b u^0_1  - 2(2ac - b^2) \ttp_0 = - 3 b u^0_1 >0,\\
\label{eq-detAd}
\det A (\bt^{\rm fold, d} (\bu_0) )
& = &
b u^0_1  -  \frac{1}{2} (2ac - b^2) \ttp_0 =0,
\eeqa
where the positive condition follows from \eq{eq-bu01}.
In other words, the image $\bt^{\rm fold, d} (\bu_0) $ has infinite
magnification and so is located on the critical curve.
On the other hand, the image 
$\bt^{\rm fold, s} (\bu_0) $ is finitely magnified
and has positive parity:
\beq
\label{eq-mag-on-s}
\mulons (\bu_0) \equiv \mul (\bt^{\rm fold, s} (\bu_0))=  \frac{1}{3|b u_1^0|}
= \frac{-1}{3b u_1^0},
\eeq 
where the rightmost equality follows from \eq{eq-bu01}.
Now, the root $\th^{\rm fold,s}_2 (\bu_0)$ determines the
caustic point $\bu^{\rm fold,s}_c = (u^{\rm fold,s}_{c,1}, u^{\rm fold,s}_{c,2}),$
where
\beq
u^{\rm fold,s}_{c,1} = \frac{3(b^2-2ac)}{2b} [\th^{\rm fold,s}_2 (\bu_0)]^2,
\qquad
u^{\rm fold,s}_{c,2} = \frac{b^2-2ac}{c} [\th^{\rm fold,s}_2(\bu_0)]^3.
\eeq
By \eq{eq-th2fsd}, we have  
\beq
u^{\rm fold,s}_{c,1}        
= 4 u^0_1.
\eeq
Consequently,  the horizontal distance from the source at
the fold point $\bu_0$ to the caustic point
$\bu^{\rm fold,s}_c$ is nonzero:
\beq
\label{eq-hsfs-n}
\hsfs \equiv  |u^0_1  -  u^{\rm fold,s}_{c,1}| = 3 |u^0_1| \neq 0.
\eeq
By \eq{eq-mag-hd} or (\ref{eq-detAs}), the magnification of $\bt^{\rm fold, s} (\bu_0) $
can be expressed as
\beq
\label{eq-mag-on}
\mulons (\bu_0) =  \frac{1}{|b| \hsfs}.
\eeq
In other words, the magnification of  $\bt^{\rm fold,s} (\bu_0)$
is not inversely proportional to the horizontal (or ordinary) distance 
from $\bu_0$ to
the caustic (which is zero since the source sits on the caustic), 
rather from  $\bu_0$ 
to the caustic point $\bu^{\rm fold,s}_c$.
Analogous to $\th^{\rm fold,s}_2 (\bu_0)$, 
the root  $\th^{\rm fold,d}_2 (\bu_0)$ in \eq{eq-th2fsd}
yields a caustic point
$\bu^{\rm fold,d}_c = (u^{\rm fold,d}_{c,1}, u^{\rm fold,d}_{c,2})$.
Since
\beq
u^{\rm fold,d}_{c,1} = \frac{3(b^2-2ac)}{2b} [\th^{\rm fold,d}_2 (\bu_0)]^2 = u^0_1,
\eeq
the horizontal distance from $\bu_0$ to 
$\bu^{\rm fold,d}_c$ is zero,
\beq
\hsfd = |u^0_1  -  u^{\rm fold,d}_{c,1}| = 0.
\eeq
Since $\mulond (\bu_0) =  1/(|b| \hsfd)$, the vanishing of the horizontal distance 
gives the earlier result that  
$\bt^{\rm fold,d} (\bu_0)$ is infinitely magnified.

As $\bu$ approaches a fold point $\bu_0$
from outside the caustic, we have
$D(\bu) \rightarrow 0$ and
$ \th^{\rm out}_2 (\bu) \rightarrow \th^{\rm fold, s}_2 (\bu_0)$
(using \Eq{eq-x2out}).  This implies that the outside local image centroid
obeys 
\beq
\label{eq-outtofolds}
\btclc^{\rm out} (\bu) \rightarrow \bt^{\rm fold, s} (\bu_0).
\eeq
However, when 
$\bu$ reaches $\bu_0$, the centroid $\btclc^{\rm out} (\bu_0)$
not only meets $\bt^{\rm fold, s} (\bu_0)$, 
but also the infinitely magnified image $\bt^{\rm fold, d} (\bu_0)$.
The latter image suddenly appears when $\bu = \bu_0$  
and dominates $\bt^{\rm fold, s} (\bu_0)$ in magnification.
This causes
the image centroid 
to have a double-value (see \Eq{eq-ic-double} below) and creates a discontinuous jump 
from $\bt^{\rm fold, s} (\bu_0)$ to $\bt^{\rm fold, d} (\bu_0)$.
Further discussion of this jump discontinuity will be given below
(see equation Section~\ref{subsec-jump} and
Figure~\ref{fig:example2}).

\subsubsection{Local Case: Source Inside the Cusp\label{sec:interior}}

For sources inside and near the cusp
(i.e., $D(\bu) < 0$), \eq{eq-cubicpol} has
three real simple roots:
\beqa
&& \th^{\rm in, 1}_2 (\bu) = 2 \sqrt{ \frac{-\ttp}{3} } 
\cos\left[\frac{\vartheta (\bu)}{3}\right], \hspace{.5in}  
\th^{\rm in, 2}_2 (\bu) = - 2 \sqrt{ \frac{-\ttp}{3} } 
\cos\left[\frac{\vartheta(\bu) - \pi}{3}\right], 
\nonumber \\
\nonumber \\
&& \th^{\rm in, 3}_2 (\bu) = -  2 \sqrt{ \frac{-\ttp}{3} } 
\cos\left[\frac{\vartheta (\bu) + \pi}{3}\right],
\nonumber \\
\label{eq-in}
\eeqa
where
\beq
\vartheta (\bu) = \cos^{-1} \left[\frac{ (-\ttq/2)}{\sqrt{ (- \ttp/3)^3}  }\right].
\eeq
Since $D(\bu) <0$, it follows that
$(\ttp/3)^3 < - (\ttq/2)^2 \le 0$, which
yields $\ttp <0$ and  $b u_1 <0$ (since
$2 ac - b^2 >0$).   In addition,
we have $|\ttq/2|/\sqrt{ (- \ttp/3)^3} <1$, or, 
$- 1 < (-\ttq/2)/\sqrt{ (- \ttp/3)^3} <1$.
Consequently, 
\beq
 0 < \vartheta (\bu) < \pi
\eeq
for $\bu$ inside the caustic.  The values
$\vartheta (\bu) =0$ and $\vartheta (\bu) = \pi$
occur when $\bu$ is on the top and bottom fold arcs, respectively,
that abut the cusp (see discussion below).
The value $\vartheta (\bu) = \pi/2$ is for $\bu$ on the
$u_1$-axis inside the caustic.

A source inside and near the cusp
has three local images determined by \eq{eq-in}:  
\beqa
\label{eq-images-in}
\bt^{\rm in, 1} (\bu) 
& = & \left(\frac{u_1}{c}  + 
\frac{2b}{3c}\ttp \cos^2 \left[\frac{\vartheta (\bu)}{3}\right],
   2 \sqrt{ \frac{-\ttp}{3} } 
\cos\left[\frac{\vartheta(\bu) }{3}\right]\right),\nonumber \\
\bt^{\rm in, 2} (\bu) 
& = & \left(\frac{u_1}{c}  + 
\frac{2b}{3c}\ttp\cos^2 \left[\frac{\vartheta (\bu) - \pi}{3}\right],
    - 2 \sqrt{ \frac{-\ttp}{3} }
\cos\left[\frac{\vartheta (\bu) -  \pi}{3}\right]\right),\nonumber \\
\bt^{\rm in, 3} (\bu)
& = & \left(\frac{u_1}{c}  +
\frac{2b}{3c}\ttp\cos^2 \left[\frac{\vartheta (\bu) + \pi}{3}\right],
  - 2 \sqrt{ \frac{-\ttp}{3} } 
\cos\left[\frac{\vartheta (\bu) + \pi}{3}\right]\right).\nonumber\\
\label{x123}
\eeqa
We saw that the condition $D(\bu) <0$ yields 
$b u_1 <0 $ and
$0 <  \vartheta (\bu) < \pi$
for sources inside the caustic. The latter,
implies
\beqa
\det A (\bt^{\rm in, 1} (\bu) )
& = &
b u_1\left[1 - 4  \cos^2 \left( \frac{\vartheta  (\bu)}{3} \right)\right] >0,\nonumber \\
\det A (\bt^{\rm in, 2}  (\bu))
& = &
b u_1\left[1 - 4  \cos^2 \left( \frac{\vartheta  (\bu) - \pi}{3} \right)\right] >0,\nonumber \\
\det A (\bt^{\rm in, 3}  (\bu))
& = &
b u_1\left[1 - 4  \cos^2 \left( \frac{\vartheta  (\bu) + \pi}{3} \right)\right] <0.\nonumber \\
\eeqa
In other words,
the lensed images
$\bt^{\rm in, 1}$ and $\bt^{\rm in, 2}$  have positive parity, while
$\bt^{\rm in, 3}$
has negative parity.  It is useful to note that
\beqa
\det A (\bt^{\rm in, 1} (\bu) )
& = &0,~{\rm for}~\theta=\pi, \nonumber \\
\det A (\bt^{\rm in, 2}  (\bu))
& = &0,~{\rm for}~\theta=0,\nonumber\\
\det A (\bt^{\rm in, 3}  (\bu))
& = &0,~{\rm for}~\theta=0,\pi.\nonumber\\
\eeqa

\begin{figure}
\plotone{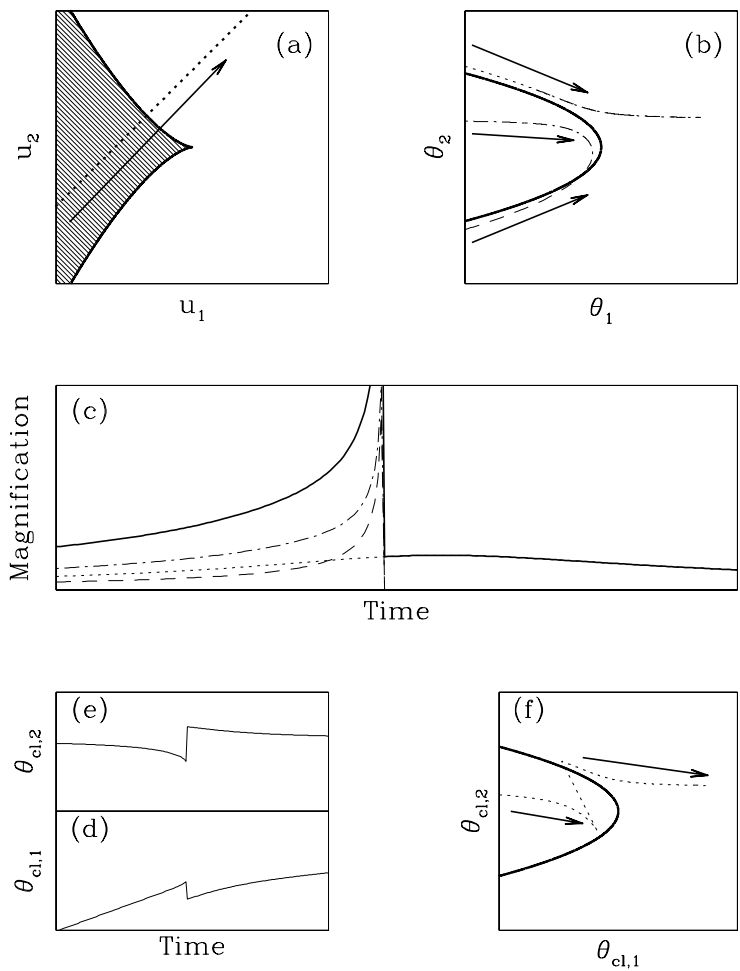}
\caption{
Local gravitational lensing for source passing
near a cusp; same as Figure \ref{fig:example1}, except for a trajectory which crosses 
one fold caustic. The dotted lines in (b) show the image path for the
portion of the source trajectory in (a) outside the caustic,
while the dashed and dashed-dot lines are the paths for the other
two local images.  Panel (c) depicts  the magnifications of the three
local images in (b); the solid line is the total magnification of the
three images.  The dotted path in (f) is the centroid trajectory for the images.}
\label{fig:example2}
\end{figure}

\begin{figure}
\plotone{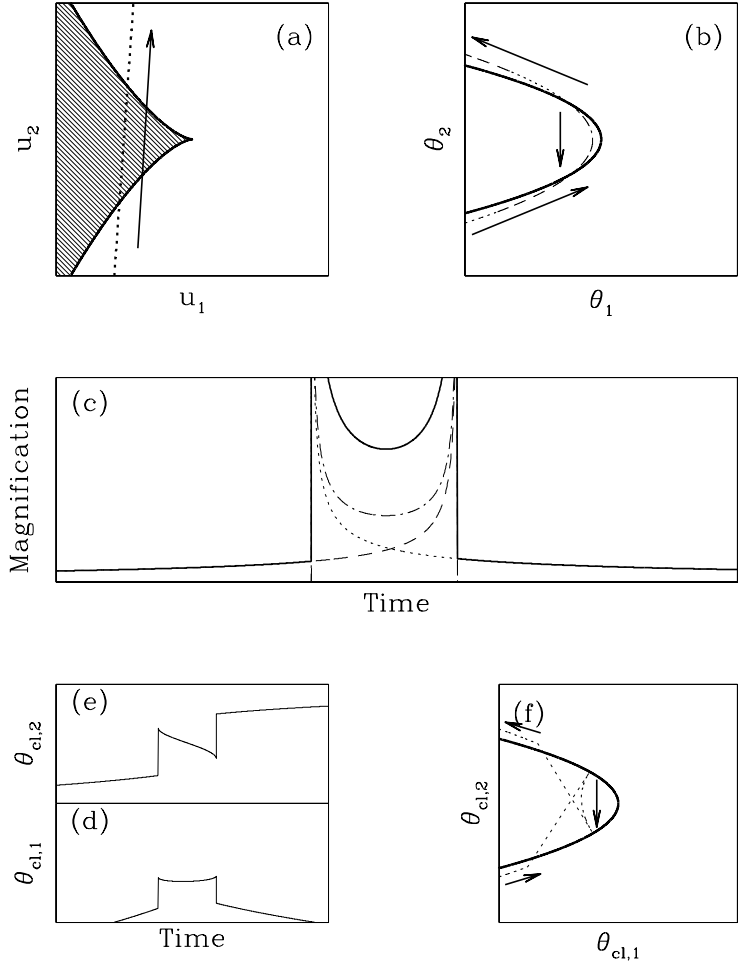}
\caption{ 
Same as Figure \ref{fig:example1} and \ref{fig:example2}, except for a trajectory which crosses 
both upper and lower fold caustics.}
\label{fig:example3}
\end{figure}

Let us now consider the behavior of three images  in \eq{eq-images-in}
near the caustic --- see Figures~\ref{fig:example2} and \ref{fig:example3}.
Equation (\ref{eq-images-in})  shows that as $\bu \rightarrow \bmo$ (source
approaches the cusp) from inside the caustic, we obtain
\beq
\label{eq-3images-origin}
\bt^{\rm in, \ell} (\bu) \rightarrow \bmo, \qquad  \ell = 1,2,3,
\eeq
i.e., all three images merge at the point on the critical curve that
maps to the cusp. We now investigate the image behavior as the
source approaches a fold caustic.
There are two fold arcs abutting the cusp, one in lying above the $u_1$-axis
and the other below.   If the source is on one of these fold arcs, then
$D (\bu) =0$ and $\ttp \ttq \neq 0$.
At a point $\bu_0$ on the fold arc above the $u_1$-axis,
we have $ (-\ttq_0/2) = \sqrt{(-\ttp_0/3)^3}$, which yields
$\vartheta (\bu_0) =0$.  By \eq{x123}, as the source position
$\bu$ approaches $\bu_0$ from the interior of the cusped caustic,
we have
\beq
\label{eq-in-x1s-2d3-fold}
\bt^{\rm in, 1} (\bu) \longrightarrow \bt^{\rm fold, s} (\bu_0),
\hspace{.3in}
\bt^{\rm in, 2} (\bu) \longrightarrow \bt^{\rm fold, d} (\bu_0),
\hspace{.3in}
\bt^{\rm in, 3} (\bu) \longrightarrow \bt^{\rm fold, d} (\bu_0).
\eeq
If the source crosses the fold transversely through the point $\bu_0$ and
continues on to outside the fold, then
the lensed image moving along $\bt^{\rm in, 1} (\bu)$ passes through
$\bt^{\rm fold, s} (\bu_0)$
and continues along $\bt^{\rm out} (\bu)$. In addition,
the other two images moving along $\bt^{\rm in, 2} (\bu)$ and
$\bt^{\rm in, 3} (\bu)$ merge into a single lensed image 
at $\bt^{\rm fold, d} (\bu_0)$,
and then disappear as the source moves outside the caustic.
These results are depicted in
Figure~\ref{fig:example2}(b).  Similar results
occur if the source travels from outside the caustic approaching 
a point $\bu_0$ 
lying on the fold arc below the $u_1$-axis,
see Figure~\ref{fig:example3}(b).
In this case,
$ (-\ttq_0/2) = -\sqrt{(-\ttp_0/3)^3}$, which yields
$\vartheta (\bu_0) = \pi$.  
As the source approaches $\bu_0$ from outside, there is one 
image locally and the image moves along $\bt^{\rm out}$.
Equation (\ref{x123}) shows that as $\bu \rightarrow \bu_0$
from outside the caustic, we have
\beq
\label{eq-from-out}
\bt^{\rm out} (\bu) \longrightarrow \bt^{\rm fold, s} (\bu_0)
= \bt^{\rm in, 2} (\bu_0).
\eeq 
When the source reaches $\bu_0$, an infinitely magnified lensed
image $\bt^{\rm fold, d} (\bu_0)$ suddenly appears.  As
the source moves transversely through $\bu_0$ and enters inside
the caustic, \eq{x123} yields that
$\bt^{\rm fold, d} (\bu_0)$ splits into two images
that move along $\bt^{\rm in, 1} (\bu)$ and
 $\bt^{\rm in, 3} (\bu)$.
Equivalently, if a source at $\bu$ moves from inside the caustic to a 
point $\bu_0$ on the fold arc below the $u_1$-axis, we have
that $\bt^{\rm in, 1} (\bu)$ and
$\bt^{\rm in, 3} (\bu)$ merge at $\bt^{\rm fold, d} (\bu_0)$:
\beq
\label{eq-in-x2s-1d3-fold}
\bt^{\rm in, 2} (\bu) \longrightarrow \bt^{\rm fold, s} (\bu_0),
\hspace{.1in}
\bt^{\rm in, 1} (\bu) \longrightarrow \bt^{\rm fold, d} (\bu_0),
\hspace{.1in}
\bt^{\rm in, 3} (\bu) \longrightarrow \bt^{\rm fold, d} (\bu_0).
\eeq

The magnifications of the images $\bt^{\rm in, \ell} (\bu)$
are
\beq
\label{eq-mag-in-123}
\mu^{\rm in}_\ell (\bu)
\equiv \mu (\bt^{\rm in, \ell} (\bu) ) = \frac{1}{|b| \hsinell}, \qquad \ell =1,2,3,
\eeq
where the horizontal distances are given by
\beqa
\label{eq-hs-in-123}
&& \hsina 
 = 
|u_1\left(1 - 4 \cos^2 [\vartheta (\bu)/3]\right)|,
\qquad 
\hsinb
 = 
|u_1\left(1 - 4 \cos^2 [
   (\vartheta (\bu) - \pi)/3] \right)|,
\nonumber \\
&& \hsinc
 = 
|u_1\left(1 -  4 \cos^2 [( \vartheta (\bu) + \pi)/3] \right)|.
\nonumber \\
\eeqa
Observe that if the source is inside the caustic, but on the $u_1$-axis,
then $\vartheta (u_1, 0) =  \pi/2$, which yields
\beq
\label{eq-magin-u1}
\mu^{\rm in}_1 (u_1,0) = \frac{1}{2|b u_1|} = \frac{-1}{2b u_1},
\qquad \mu^{\rm in}_2 (u_1,0) = \frac{1}{2|b u_1|} = \frac{-1}{2b u_1},
\qquad 
\mu^{\rm in}_3 (u_1,0) = \frac{1}{|b u_1|} = \frac{-1}{b u_1}.
\eeq
In other words, the magnification is inversely proportional to
the distance of the source (on the $u_1$-axis) from the cusp.
The magnification of the images are
shown in Figures~\ref{fig:example2}(b) and \ref{fig:example3}(b)
for a source in rectilinear motion.
It follows from \eq{eq-magin-u1} that
\beq
\mu^{\rm in}_3 (u_1,0) =  \mu^{\rm in}_1 (u_1,0) + \mu^{\rm in}_2 (u_1,0).
\eeq
In fact, this relationship holds for any position $\bu$ inside and
near the cusp 
(e.g., \citealt{sw92, zak95, zak99}):
\beq \mu^{\rm in}_3 (\bu) =  \mu^{\rm in}_1 (\bu) + \mu^{\rm in}_2 (\bu),\eeq
where $D (\bu) <0$.
Hence, the total magnification of the three local images is 
\beq
\label{eq-mag-in}
\mulin (\bu) = 
\mu^{\rm in}_1 (\bu)  + \mu^{\rm in}_2 (\bu) + \mu^{\rm in}_3 (\bu)
=
2 \left[\mu^{\rm in}_1 (\bu)  + \mu^{\rm in}_2 (\bu) \right].
\eeq

The image centroid  of the three images for source positions
inside and near the cusp
is given by
\beq
\label{eq-col-in-2}
\btclc^{\rm in}  =
\frac{\mu^{\rm in}_1 \bt^{\rm in,1} +
\mu^{\rm in}_2 \bt^{\rm in,2} +
\mu^{\rm in}_3 \bt^{\rm in,3} }
{\mu^{\rm in}_1 + \mu^{\rm in}_2 + \mu^{\rm in}_3}
=
\frac{1}{2} \left[\btclc^{\rm in (1,2)} + \bt^{\rm in,3} \right],
\eeq
where
$\btclc^{\rm in (1,2)}$ is the centroid
of the images $\bt^{\rm in, 1}$ and $\bt^{\rm in, 2}$, i.e.,
\beq
\label{eq-col-in12}
\btclc^{\rm in (1,2)} =
\frac{
\mu^{\rm in, 1}  \bt^{\rm in, 1}
  +   \mu^{\rm in, 2} \bt^{\rm in, 2}}{
\mu^{\rm in, 1} + \mu^{\rm in, 2}}.
\eeq
The image centroid for a source in rectilinear motion is shown
in Figures~\ref{fig:example2}(f) and \ref{fig:example3}(f).

Equation (\ref{eq-in-x1s-2d3-fold}) shows that as
$\bu \rightarrow \bu_0$ from inside the caustic,
where $\bu_0$ lies on the fold arc above the $u_1$-axis,
we obtain
$\mu^{\rm in, 1} (\bu) \rightarrow \mu^{\rm fold, s}_{\rm loc} (\bu_0),
$
where $0 < \mu^{\rm fold, s}_{\rm loc} (\bu_0) < \infty$,
and
$\mu^{\rm in, 2}(\bu) \rightarrow \infty$.
This yields
\beq
\frac{\mu ^{\rm in, 1}(\bu)}{
\mu^{\rm in,1}(\bu) + \mu^{\rm in,2}(\bu)}
\rightarrow 0, \qquad
\frac{\mu^{\rm in, 2}(\bu)}{
\mu^{\rm in,1}(\bu) + \mu^{\rm in,2}(\bu)}
\rightarrow 1 \qquad \quad \mbox{as} \ \bu \rightarrow \bu_0.
\eeq
Similarly, \eq{eq-in-x2s-1d3-fold}
shows that if $\bu_0$ is on the fold arc below the
$u_1$-axis, then $\mu^{\rm in, 2}(\bu) \rightarrow \mu^{\rm fold, s}_{\rm loc}(\bu_0)
$
and
$\mu^{\rm in, 1}(\bu) \rightarrow \infty$.  Consequently,
\beq
\frac{\mu^{\rm in, 1}(\bu)}{
\mu^{\rm in,1}(\bu) + \mu^{\rm in,2}(\bu)}
\rightarrow 1, \qquad
\frac{\mu^{\rm in, 2}(\bu)}{
\mu^{\rm in,1}(\bu) + \mu^{\rm in,2}(\bu)}
\rightarrow 0 \qquad \quad \mbox{as} \ \bu \rightarrow \bu_0.
\eeq
It follows that
as $\bu \rightarrow \bu_0$ from inside the caustic,
we have $\btclc^{\rm in (1,2)} (\bu) \rightarrow
\bt^{\rm fold,d}$ and $\bt^{\rm in, 3}(\bu) \rightarrow
\bt^{\rm fold,d}$.   Hence,
\beq
\label{eq-colfold}
\btclc^{\rm in} (\bu)
\rightarrow \bt^{\rm fold, d} (\bu_0).
\eeq
as $\bu \rightarrow \bu_0$ from inside the caustic.
Equations (\ref{eq-outtofolds}) and (\ref{eq-colfold}) then show
that for $\bu = \bu_0$, the local image centroid is double valued
and so has a jump discontinuity
(Figures~\ref{fig:example2}(f) and \ref{fig:example3}(f)):
\beq
\label{eq-ic-double}
\btclc (\bu_0) = \{ \bt^{\rm fold, s} (\bu_0), \ \bt^{\rm fold, d} (\bu_0) \}.
\eeq

\subsubsection{Local Case: Summary of  Local Results\label{sec:sumloc}}

The magnification of the 
local images associated with a source
at angular vector position $\bu$ 
near the cusp can now be summarized as follows:
\beq
\label{eq-loc-mag}
\mul (\bu)
=
\left\{
\begin{array}{ll}
\mulout (\bu) &  \quad \mbox{$\bu$ outside caustic},\\ \\
\infty , & \quad  \mbox{$\bu$ on caustic},\\ \\
2 \left[\mu^{\rm in}_1 (\bu)  + \mu^{\rm in}_2 (\bu) \right] &
\quad \mbox{$\bu$ inside caustic},
\end{array}
\right.
\eeq
where the magnifications $\mulout$ and $\mu^{\rm in}_\ell$, $\ell = 1,2$, 
 are given by equations
(\ref{eq-mag-out}) and (\ref{eq-mag-in-123}), respectively.  
These magnifications are shown as a function of source position
in Figures~\ref{fig:example1}-\ref{fig:conmat}.  
The image centroid is given by:
\beq
\label{eq-loc-centroid}
\btclc (\bu)
= 
\left\{
\begin{array}{ll}
\bt^{\rm out} (\bu) &  \quad \mbox{$\bu$ outside caustic},\\ \\
\left\{\btclc^{\rm fold, d} (\bu_0),\  \btclc^{\rm fold, s} (\bu_0)\right\}, &
\quad  \mbox{$\bu = \bu_0$ on fold arc},\\ \\
\bmo, &\quad  \mbox{$\bu = \bmo$ on cusp},\\ \\
(1/2) \left[\btclc^{\rm in (1,2)} (\bu)  + \bt^{\rm in,3} (\bu) \right]&  
\quad \mbox{$\bu$ inside caustic},
\end{array}
\right.
\eeq
where the angular vector $\bt^{\rm out}$ is given by \eq{eq-out-li},
$\btclc^{\rm fold, d}$ by \eq{eq-li-foldd}, $\btclc^{\rm fold, s}$ by
\eq{eq-li-folds}, $\bt^{\rm in,3}$ by \eq{eq-images-in}, and
$\btclc^{\rm in (1,2)}$ by \eq{eq-col-in12}.  Note that the above
equations can be applied to a general (not necessarily rectilinear)
source trajectory $\bu = \bu (t)$.  The local centroid is illustrated
in Figures~\ref{fig:example1}(f), \ref{fig:example2}(f), and
\ref{fig:example3}(f) for rectilinear source motion.

\subsection{Global Case: Point Sources Near a Cusp\label{sec:global}}

\subsubsection{Magnification and Centroid of All Images}

Suppose that the source moves rectilinearly, i.e., its trajectory
is given by
\beq
\label{eq-trajectory}
\bu (t) = \bu_c +  (t - t_c) \dot{\bu},
\eeq
where $\bu_c$ is the position at which the source intersects
at time $t=t_c$ the tangent line to the cusp,  which coincides with the 
the $u_1$-axis, and $\dot{\bu}$ is the constant angular velocity vector
of the source,
\beq
\dot{\bu}= \left(
\frac{\cos \phi }{\te}, \frac{\sin \phi}{\te}\right).
\eeq
Recall that $\te= \dol \thetae/\rv_\perp$, where 
$\thetae$ is the angular Einstein radius and $\rv_\perp$ is the proper transverse
speed of the lens relative to the observer-source line-of-sight.
Here $\phi$ is the angle of the source's trajectory with respect to 
the $u_1$-axis (i.e., tangent line to the cusp).

As the source moves along the rectilinear path
in \eq{eq-trajectory},
the angular positions of the images unaffiliated with the cusp
follow continuous trajectories and their magnifications
are positive, finite,  and continuous.
Moreover, we assume that for source trajectories near the cusp,
the image magnifications and positions of the unaffiliated
images are slowly varying functions of the source position $\bu$.
Denote the centroid and total magnification of the unaffiliated images
by $\bt_0$ and $\mu_0$, respectively.  Then the magnification $\mut$ 
and image centroid $\btcl$ 
of all the images are given by
\beq
\label{eq-mag-tot}
\mut (\bu) = \mul (\bu) + \mu_0 (\bu) 
\eeq
and
\beq
\label{eq-global-cl}
\btcl (\bu) = \frac{1}{\mu_{tot} (\bu)} \left[\mul (\bu) \btclc (\bu) +
\mu_0 (\bu) \bt_0 (\bu)\right].
\eeq
The photometric and astrometric observables $\mut$ and $\btclc$, respectively,
can be expressed as functions of time by 
replacing $\bu$ in equations (\ref{eq-mag-tot}) and (\ref{eq-global-cl}) by
the rectilinear source trajectory $\bu (t)$ in \eq{eq-trajectory}, and
Taylor expanding $\bt_{0}$ and $\mu_0$ to first order  about the crossing time 
$t = \tcc$ of the 
tangent line to the cusp.
This yields: 
\beq
\label{eq-mag-tot-time}
\mut (t)  =  \mul (t) + \mu_0(t),
\eeq
where
$\mul (t) = \mul (\bu(t))$  and 
$\mu_0 (t)  =  \mu_{0,c} + (t-\tcc) \dot{\mu}_{(0,c)}$ 
with $\dot{\mu}_{(0,c)} = {\rm d}\mu_0 /{\rm d}t|_{t=t_c},$
and
\beq
\label{eq-global-cl-time}
\btcl (t)  =  \frac{1}{\mu_{tot} (t)} \left[\mul (t) \btclc (t) + 
\mu_0(t)\bt_0(t)\right],
\eeq
where $\btclc (t) = \btclc (\bu(t))$ and
$\bt_0   = 
\bt_{0,c} + (t-\tcc) \dot{\bt}_{(0,c)}
$
with $\dot{\bt}_{(0,c)} = {\rm d}\bt_0 /{\rm d}t|_{t=t_c}.$

\subsubsection{Image Centroid Jump-Discontinuity at Fold Crossing}
\label{subsec-jump}

Assume that a source in rectilinear motion 
$\bu (t)$  transversely crosses the point $\bu_0$ on one of the fold arcs
abutting the cusp.
As $\bu (t) \rightarrow \bu_0$ from inside
(respectively, outside) the cusp, denote the limiting value
of the image centroid
$\btcl (\bu (t))$ by $\btcl^+ (\bu_0)$
(respectively, $\btcl^- (\bu_0)$).
The jump-discontinuity vector at $\bu_0$
for the image centroid and shifted image centroid are the same:
\beq
\btcl^{\rm jump} (\bu_0)
= \btcl^+ (\bu_0) - \btcl^- (\bu_0)
= \bTcl^{\rm jump} (\bu_0).
\eeq
As $\bu \rightarrow \bu_0$ from inside the cusp, we have
\beq
\frac{\mu_0 (\bu)}{\mut (\bu)} \rightarrow \frac{1}{1 + \mul (\bu_0)/\mu_0 (\bu_0)} = 0,
\eeq
(since $\mul (\bu) \rightarrow \infty$ and $\mu_0 (\bu)$ is generally
non-divergent).   Therefore, from \eq{eq-global-cl}, it follows
that $\btcl^+ (\bu_0) =  \bt^{\rm fold, d} (\bu_0)$, i.e.\ the contribution
from the images not associated with the cusp is negligible. 
In contrast, as $\bu \rightarrow \bu_0$ from {\it outside} the cusp, we have
\beq
\frac{\mu_0 (\bu)}{\mut (\bu)} \rightarrow \frac{1}{1 + \mu_{\rm loc}^{\rm out} (\bu_0)/\mu_0 (\bu_0)} \ne 0,
\eeq
where the last inequality holds provided that $\bu_0\ne {\bf 0}$.   Thus,
we generally have that $\btcl^- (\bu_0) \ne \bt^{\rm fold, s} (\bu_0)$, and
the contribution from images not associated with the cusp can be substantial.  
However, due to the fact that the magnification of the image exterior to the cusp
is $\gg 1$ for positions close to the cusp, it likely a fair approximation
to ignore the contribution from the other images.  We will therefore
assume that $\mu_{\rm loc}^{\rm out} (\bu_0)/\mu_0 (\bu_0)\gg 1$, and
thus $\btcl^- (\bu_0) \simeq \bt^{\rm fold, s} (\bu_0)$.  Consequently,
\beq
\label{eq-jump-1}
\btcl^{\rm jump} (\bu_0)
=  \bt^{\rm fold, d} (\bu_0)
    - \bt^{\rm fold, s} (\bu_0)
\eeq
Since $ \bt^{\rm fold, d} (\bu_0) \neq \bt^{\rm fold, s} (\bu_0)$,
we obtain a discontinuous jump in the image centroid as the source
passes through the fold point $\bu_0$.
There is {\it no} discontinuous jump if the source passes through the
cusp (since $\btcl^{\rm jump} (\bmo) = \bmo$).
Since by \eq{eq-discriminant} we have 
$(-\ttq_0/2)^{1/3} = \mp (-\ttp_0/3)^{1/2}$, where the minus (respectively, plus)
corresponds to $\bu_0$ on the fold arc below (respectively, above) the $u_1$-axis,
equations (\ref{eq-li-foldd}) and (\ref{eq-li-folds}) yield 
\beq
\label{eq-jump-2}
\btcl^{\rm jump} (\bu_0)
 =  
\left(-\frac{b}{2 c} \ttp_0, \mp 3 \sqrt{\frac{-\ttp_0}{3}} \right)
=
\left(\frac{b^2 u^0_1}{c (2ac-b^2)}, 
    \mp \sqrt{ \frac{-6 b u^0_1}{2ac-b^2} } \right), 
\eeq
where as before the $\mp$ signs are for $\bu_0$ on the respective
fold arcs below and above the tangent line to the cusp
(i.e., the $u_1$-axis).
Note that $b u^0_1 < 0$ by \eq{eq-bu01}.

For $\bu_0$ sufficiently close to  the cusp at $\bmo$, the variable $u^0_1$ dominates
$(u^0_1)^2$.   
We can then approximate the magnitude
of the jump as follows: 
\beq
\label{eq-jump-mag}
\left| \btcl^{\rm jump} (\bu_0) \right|
= 
\left[ \left(\frac{b^2  u^0_1}{c (2ac-b^2)}\right)^2 + 
\frac{-6 b u^0_1}{2ac - b^2} \right]^{1/2}
\simeq \sqrt{\frac{-6 b  u^0_1}{2ac - b^2}}.
\eeq
Therefore, the magnitude of the centroid jump increases as $|u^0_1|^{1/2}$
for $|u^0_1| \ll 1$, where $|u^0_1|$ is the horizontal distance of the 
caustic point $\bu_0$ from the cusp.
Note that no jump discontinuity occurs for a source passing through the cusp.

\subsection{Extended Sources Near a Cusp\label{sec:fsources}}

\subsubsection{Magnification and Image Centroid of Extended Sources\label{sec:genfs}}

Consider an extended source with surface brightness
$S(\bu)$.  The  
average surface brightness 
is 
$\bar{S} \equiv (\pi \rho^2_*)^{-1}\ \int_\cD {\rm d}\bu \ S(\bu)$,
where $\cD$ is the disc-shaped region of the source and
$\rho_*\equiv\theta_*/\thetae$ is the angular source radius
$\theta_*$ in units of $\thetae$.  
The magnification of the finite source is then
\beq
\label{eqn:fsgen}
\mu^{\rm fs}  =  {{\int_\cD {\rm d} \bu S(\bu)   \mu(\bu)}\over{\int_\cD {\rm d} \bu S(\bu)}},
\eeq
Introduce new coordinates by
\beq
\bu' = \frac{\bu - \bucn}{\rho_*},
\eeq
where $\bucn$ is the center of the source.  Note that $|u'| \le 1$.  Then
\eq{eqn:fsgen} simplifies to 
\beq
\mu^{\rm fs}  =  \frac{1}{\pi} \int_D {\rm d}\bu' \Sn(\bu') \mu(\bu'),
\label{eq-mufsscale}
\eeq
where $\Sn (\bu) = S (\bu)/ \bar{S}.$
In the case of small source sizes $\theta_* \ll \thetae$, the magnification $\mu_0 (\bu)$
of the images not associated with the cusp is a slowly varying
function of $\bu$ over the source, which yields
\beq
\mu_{0}^{\rm fs}\simeq\mu_0(\bu_{cn})\equiv\mu_{0,{cn}}.
\eeq
The total finite source magnification becomes
\beq
\mtfs =   \frac{1}{\pi}\int_\cD {\rm d}\bu' \Sn (\bu') \mut (\bu')
= \mulfs + \mu_{0,cn},
\label{eq-mtfs}
\eeq
where 
\beq
\label{eq-mlocfs}
\mulfs  =  \frac{1}{\pi} \int_\cD {\rm d}\bu'  \Sn (\bu') \mul (\bu').
\eeq

For an extended source,  the image centroid is given by
\beq
\label{eq-ic-1}
\btclfs \equiv 
    \frac{ \int_\cD {\rm d} \bu  S(\bu) \mut (\bu) \btcl (\bu)}
{ \int_\cD {\rm d} \bu S(\bu)\mut (\bu) }
=
\frac{ \int_\cD {\rm d} \bu' \Sn (\bu') \mut (\bu')   \btcl (\bu')}
          { \pi \mtfs },
\eeq
where the denominator of the middle ratio is 
$\pi  \rho_*^2 \bar{S} \mtfs$.
For a source with angular radius $\theta_* \ll \thetae$, the centroid $\bt_0 (\bu)$
of the images not associated with the cusp varies slowly
over the source.  Consequently, since
$\mu_0 (\bu)$ also varies slowly over the
source, we get  
\beq
\mu^{\rm fs}_{0,{cn}} \bt^{\rm fs}_{0, {\rm cn}}\simeq \mu_0 (\bu_{cn}) \bt_0 (\bu_{cn})=\mu_{0,{cn}} \bt_{0, {\rm cn}},
\eeq
where we have defined $\bt_{0, {\rm cn}} \equiv  \bt_0 (\bu_{cn})$.
Equation (\ref{eq-global-cl}) then yields
\beq
\label{eq-ic-2}
\btclfs = \frac{ \mulfs}{\mtfs} \btclcfs   +    
    \frac{ \mu_{0, {\rm cn}}}{\mtfs} \bt_{0, {\rm cn}},
\eeq
where
the contribution from the images affiliated with 
the cusp  is 
\beq
\label{eq-ic-loc}
\btclcfs  
\equiv 
    \frac{ \int_\cD {\rm d} \bu  S(\bu) \mul (\bu)  \btclc (\bu)}
{ \int_\cD {\rm d} \bu  S(\bu) \mul (\bu) }
=
    \frac{ \int_\cD {\rm d} \bu' \Sn (\bu') \mul (\bu') \btclc (\bu')}
          { \pi\mulfs }.
\eeq
and from the images unrelated to the cusp is
$\bt_{0, {\rm cn}}$.

\subsubsection{Small Finite Sources on the Axes}

We would like to be able to apply the formulae in the previous
section (\S\ref{sec:genfs}) to the analytic expressions
derived for the sources near a cusp (see \S\ref{sec:sumloc}),
to find analytic (or semi-analytic) 
expressions for the finite-source magnification and image centroid of
arbitrarily large sources
with arbitrary positions with respect to the cusp.  Unfortunately, due to the complicated 
forms for these quantities for arbitrary point source positions, this
is effectively impossible, and numerical methods must be employed.  We take
this approach in \S\ref{sec:mb9728}.   However, it is possible
to obtain analytic expressions under certain restrictive assumptions.
Specifically, we now determine the local magnification and local image
centroids for relatively small uniform and limb-darkened extended sources on
the $u_1$-  and $u_2$-axes.  

The magnification and image centroid of an extended source are given
by \eq{eq-mtfs} and \eq{eq-ic-2}, respectively, where the key terms
are the local magnification (\Eq{eq-mlocfs}) and
local image centroid (\Eq{eq-ic-loc}).  The quantitative forms
for the magnification and centroid of a finite sized source depends
on the specific form of the surface brightness profile $S(\bu)$.  
In this section, we derive simplify expressions that involve integrals over
arbitrary surface brightness profiles.  In the next section (\S\ref{sec-us-limb}), 
we adopt specific surface brightness profiles to derive semi-analytic
expressions for the centroid and magnification for small sources on the fold
axes.  

We first consider sources on the $u_1$-axis, such that $\cD$ is
entirely exterior to the caustic, i.e., such that $C(\bu)>0$ for all $\bu$ in $\cD$. Express the center of the source as
$\bu_{cn} = (z \rho_*, 0)$, where $z \ge 1$ with $z=1$ corresponding
to the boundary of the source meeting the cusp and $z >1$ yielding a
source on the positive $u_1$-axis away from the cusp.  We shall assume
that for a sufficiently small source $\rho_* \ll 1$, points $\bu = (u_1,
u_2)$ inside the source are such that $u_2 \approx 0$ and so
\eq{eq-x2out-1} yields $\th_2 (\bu) \approx 0$.  By equations
(\ref{eq-mag-out}) and (\ref{eq-hs-out}), the local magnification at
points $\bu$ inside a sufficiently small uniform source centered at
$\bu_{cn} = (z \rho_*, 0)$ can be approximated by $\mulout (\bu) = |b
u_1|^{-1}$, or, equivalently,
\beq
\label{eq-mag-out-u1}
\mulout (\bu')  =  \frac{1}{\rho_*|b|} \frac{1}{|u'_1 + z|},
\eeq
Using \eq{eq-mag-out-u1}, the finite source local 
magnification (\Eq{eq-mlocfs}) becomes
\beq
\label{eq-mlocfs-out-2}
\muloutfs (z)
= \frac{1}{\rho_*|b|} \cP_\Sn (z),
\eeq
where
\beq
\label{eq-PSN}
\cP_\Sn (z)
= \frac{1}{\pi}
\int_{-1}^1 \rd u'_1  \frac{1}{|u'_1 + z|}
  \int_{-\sqrt{1-(u'_1)^2}}^{\sqrt{1-(u'_1)^2}}
 \rd u'_2   \Sn (u'_1, u'_2).
\eeq

Now, at points $\bu$ inside
the source, \eq{eq-out-li} and \eq{eq-colout} show
that  the associated local image centroid
can be approximated as follows (since $\th_2 (\bu) \approx 0$):
\beq
\label{eq-out-li-u1}
\btclcout (\bu) =
\left(\frac{u_1}{c}, 0 \right)
=  \btclcout (\bu_{\rm cn}) + \frac{\rho_* u'_1}{c} \ui,
\eeq
where $\ui = (1,0)$.
Inserting \eq{eq-out-li-u1} into the
middle ratio in \eq{eq-ic-loc} yields
a simple formula for the local image centroid:
\beq
\label{eq-ic-loc-out-2}
\btclcoutfs (z)
= 
\btclcout (\bu_{\rm cn})+ 
\frac{ {\ui}}{|b| c \muloutfs} \cQ_{S_N} (z),
\eeq
where
\beq
\label{eq-QSN}
\cQ_{S_N} (z) \equiv 
\frac{1}{\pi} \int_{-1}^1 \rd u'_1 u'_1 \frac{1}{|u'_1 + z|}
 \int_{-\sqrt{1-(u'_1)^2}}^{\sqrt{1-(u'_1)^2}}
 \rd u'_2  \Sn (u'_1, u'_2).
\eeq

For sources on the $u_1$-axis for which $\cD$ is
entirely interior to the caustic, i.e.,
such that $C(\bu)<0$ for all $\bu$ in $\cD$, the resulting expressions are quite similar to those for sources
exterior to the caustic.  In particular, the local magnification at
points $\bu$ inside a sufficiently small uniform source centered at
$\bu_{cn} = (-z \rho_*, 0)$ can be approximated by,
\beq
\label{eq-mag-out-u1-in}
\mulin (\bu')  =  \frac{2}{\rho_*|b|} \frac{1}{|u'_1 + z|},
\eeq
and therefore
\beq
\label{eq-mlocfs-out-2-in}
\mulinfs (z)
= \frac{2}{\rho_*|b|} \cP_\Sn (z).
\eeq
Similarly, the centroid of the three images created when the source
is interior to the cusp can be shown to be,
\beq
\label{eq-out-li-u1-in}
\btclcin (\bu) =
\left(\frac{u_1}{c}\left[1+\frac{b^2}{2(2ac-b^2)}\right], 0 \right)
=  \btclcin (\bu_{\rm cn}) + \frac{\rho_*}{c}\left[1+\frac{b^2}{2(2ac-b^2)}\right] u'_1\ui,
\eeq
which is identical to the centroid for sources exterior to the caustic
(\Eq{eq-out-li-u1}) with the exception of the term in square brackets.
The finite source image centroid is therefore,
\beq
\label{eq-ic-loc-in-2}
\btclcinfs (z)
= 
\btclcin (\bu_{\rm cn})+ 
\frac{ {2\ui}}{|b| c \mulinfs}\left[1+\frac{b^2}{2(2ac-b^2)}\right] \cQ_{S_N} (z),
\eeq

Finally, we consider the magnification of small sources on the $u_2$ axis, for
which $\cD$ is entirely exterior to the caustic, i.e., such that 
$C(\bu)>0$ for all $\bu$ in $\cD$.  For sources on the $u_2$-axis,
the magnification is 
\beq
\label{eq-mag-out-u2}
\muloutperp (\bu')  =  \left(\frac{u_{r2}}{\rho_*}\right)^{2/3} 
\frac{1}{(u_2' + \rho_*)^{2/3}}.
\eeq
The finite source local magnification becomes
\beq
\label{eq-mlocfs-out-u2}
\muloutperpfs (z)
= \left(\frac{u_{r2}}{\rho_*}\right)^{2/3} \cR_\Sn (z),
\eeq
where
\beq
\label{eq-RSN}
\cR_\Sn (z)
= \frac{1}{\pi}
\int_{-1}^1 \rd u'_1  \frac{1}{(u'_1 + z)^{2/3}}
  \int_{-\sqrt{1-(u'_1)^2}}^{\sqrt{1-(u'_1)^2}}
 \rd u'_2   \Sn (u'_1, u'_2).
\eeq
We have found similar (semi-)analytic expressions for the centroid of small
sources centered on the $u_2$-axis; however these expressions are generally
unwieldy, and thus not very useful analytically.   Therefore, for the sake
of brevity, we will not present them here.

\subsubsection{Uniform and Limb-Darkened Sources on the Axes\label{sec-us-limb}}

In this section, we will consider two forms for the surface brightness
profile.  The first, and simplest, is a uniform source, where the 
normalized surface brightness obeys 
\beq
\Sn (\bu) \equiv 1.
\eeq
Second, we consider the following form for the normalized surface brightness,

\beq
\label{eq-ld}
\Sn (\bu)=\left\{1 - \Gamma\left[1- \frac{3}{2}\left( 1-
\frac{|\bu - \bu_{\rm cn}|^2}{\rho_*^2}\right)^{1/2}\right]\right\}.
\eeq
This form, first introduced by \citet{albrow1999b}, is applicable for
microlensing in the local group, and is appropriate to a non-uniform
stellar source with limb darkening.  This profile is parameterized by
the limb-darkening coefficient $\Gamma$, the value of which is
typically dependent on wavelength.  A key feature of \eq{eq-ld} is
that no net flux is associated with the limb darkening term.  Note that
 a uniform source is simply the specific case of \eq{eq-ld} with $\Gamma=0$.

Adopting these forms for $\Sn$, we can determine the integral
functions $\cP_{\Sn}$, $\cQ_{\Sn}$, and $\cR_{\Sn}$, which dictate the
basic astrometric and photometric behavior of small finite sources on
the axes of the cusp.  The following special case of $\cP_{\Sn}(z)$ (\Eq{eq-PSN}) is
technically handy:
\beq
\label{eq-PSn}
\cP_n (z)
= \frac{1}{\pi}
\int_{-1}^1 \rd x  \frac{1}{|x + z|}
  \ \int_{-\sqrt{1-x^2}}^{\sqrt{1-x^2}}
 \rd y (n + 1)\ [(1-x^2) - y]^n, 
\eeq
where $n = k/2$ with $k > -2$ an integer.
Using the following identity (e.g., Zwillinger 1996, p. 387, Eq. 599),
\beq
\label{eq-ref-int} 
\int_0^{\rm a} ({\rm a}^2 - y^2)^n d y 
= \frac{\sqrt{\pi}}{2}\ \frac{n!}{(n + 1/2)!} {\rm a}^{2(n + 1/2)},
\eeq
\eq{eq-PSn} simplifies to 
\beq
\label{eq-PSn-2}
\cP_n (z)
= \frac{1}{\sqrt{\pi}} \frac{(n + 1)!}{(n + 1/2)!}
\int_{-1}^1 \rd x  \frac{(1 - x^2)^{n + 1/2}}{|x + z|}.
\eeq
Similarly, we can identify the special case of $\cQ_{\Sn}(z)$ (\Eq{eq-QSN}),
\beqa
\label{eq-QSn}
\cQ_n (z)
& \equiv  &\frac{1}{\pi}
\int_{-1}^1 \rd x  x  \frac{1}{\sqrt{(x + z)^2}}
  \int_{-\sqrt{1-x^2}}^{\sqrt{1-x^2}}
 \rd y  (n + 1) [(1-x^2) - y]^n, \nonumber \\
&=& 
\frac{1}{\sqrt{\pi}} \frac{(n + 1)!}{(n + 1/2)!} 
\int_{-1}^1 \rd x  x  \frac{(1 - x^2)^{n + 1/2}}{\sqrt{(x + z)^2}},
\eeqa
and  $\cR_{\Sn}(z)$ (\Eq{eq-RSN}),
\beqa
\label{eq-RSn}
\cR_n (z)
& \equiv  &\frac{1}{\pi}
\int_{-1}^1 \rd x  x  \frac{1}{(x + z)^{2/3}}
  \int_{-\sqrt{1-x^2}}^{\sqrt{1-x^2}}
 \rd y  (n + 1) [(1-x^2) - y]^n, \nonumber \\
&=& 
\frac{1}{\sqrt{\pi}} \frac{(n + 1)!}{(n + 1/2)!} 
\int_{-1}^1 \rd x x \frac{(1 - x^2)^{n + 1/2}}{(x + z)^{2/3}}.
\eeqa

For a small uniform source outside the cusp, and on the axis of the cusp
(i.e., tangent line), \eq{eq-mlocfs-out-2} shows that 
the local magnification becomes
\beq
\label{eq-mlocus}
\mulusout (z)  =  \frac{1}{\rho_* |b|} \cP_0 (z).
\eeq 
Similarly, from \eq{eq-mag-out-u1-in}, the magnification for a small
uniform source inside the cusp, and on the axis of the cusp, is
\beq
\label{eq-mlocinus}
\mulusin (z)  =  \frac{2}{\rho_* |b|} \cP_0 (z).
\eeq 
By equations (\ref{eq-ic-loc-out-2}) and (\ref{eq-mlocus}), 
the uniform source local image centroid for source exterior to
the caustic is
\beq
\label{eq-ic-loc-out-3}
\btclcusout (z)
= 
\btclcout (\bu_{\rm cn}) +
\frac{\rho_*}{c} \frac{\cQ_0 (z)}{\cP_0 (z)} \ui,
\eeq
where $\btclcout (\bu_{\rm cn}) = ( z \rho_*/c, 0)$.  For sources interior
to the caustic, the corresponding centroid is,
\beq
\label{eq-ic-loc-in-3}
\btclcusin (z)
= 
\btclcin (\bu_{\rm cn}) +
\frac{\rho_*}{c}\left[1 +\frac{b^2}{2(2ac-b^2)}\right] \frac{\cQ_0 (z)}{\cP_0 (z)} \ui.
\eeq

For  a limb darkened source,
\eq{eq-PSN} becomes
\beq
\label{eq-mlocld}
\cP_{S_N}^{\rm ld} (z) = \cP_0 (z) + \Gamma [ \cP_{1/2} (z) - \cP_0 (z) ].
\eeq
Consequently, the finite source local magnification exterior to the cusp is
\beq
\label{eq-mag-fsloc}
\mulldout (z)  =  
\frac{1}{\rho_* |b|} \left( \cP_0 (z) + \Gamma [ \cP_{1/2} (z) -
\cP_0 (z) ]\right),
\eeq 
and interior to the cusp,
\beq
\label{eq-mag-fslocin}
\mulldin (z)  =  
\frac{2}{\rho_* |b|} \left( \cP_0 (z) + \Gamma [ \cP_{1/2} (z) -
\cP_0 (z) ]\right).
\eeq 
Now, equations (\ref{eq-QSN}) and (\ref{eq-QSn}) yield that 
\beq
\label{eq-QSn-ld}
\cQ_{S_N}^{\rm ld} (z) = \cQ_0 (z) + \Gamma [ \cQ_{1/2} (z) - \cQ_0 (z)].
\eeq
Thus,  using \eq{eq-QSn-ld} we see that
the local centroid exterior to the caustic (\Eq{eq-ic-loc-out-2}) becomes
\beq
\label{eq-loc-ic-out-ld}
\btclcoutld (z)
=
\btclcout (\bu_{\rm cn}) +
\frac{\rho_*}{c}
\frac{\cQ_0 (z) + \Gamma [ \cQ_{1/2} (z) - \cQ_0 (z)]}{\cP_0 (z) + 
\Gamma [ \cP_{1/2} (z) - \cP_0 (z)]} 
\ui,
\eeq
and interior to the caustic (\Eq{eq-ic-loc-in-2}),
\beq
\label{eq-loc-ic-in-ld}
\btclcinld (z)
=
\btclcin (\bu_{\rm cn}) +
\frac{\rho_*}{c}\left[1+\frac{b^2}{2(2ac-b^2)}\right]
\frac{\cQ_0 (z) + \Gamma  [ \cQ_{1/2}(z) - \cQ_0 (z)]}{\cP_0 (z) 
+ \Gamma [ \cP_{1/2} (z) - \cP_0 (z)]} 
\ui.
\eeq
Figure \ref{fig:fpar} shows the characteristic
magnification functions $\cP_0(z)$ and $\cP_{1/2}(z)$, as well
as the centroid functions $\cQ_0(z)$ and $\cQ_{1/2}(z)$.   Also shown
is the fractional difference $(\cP_{1/2} (z) - \cP_{0} (z))/\cP_0 (z)$ 
and the absolute difference $\cQ_{1/2} (z) -\cQ_0 (z)$.

\begin{figure}
\plotone{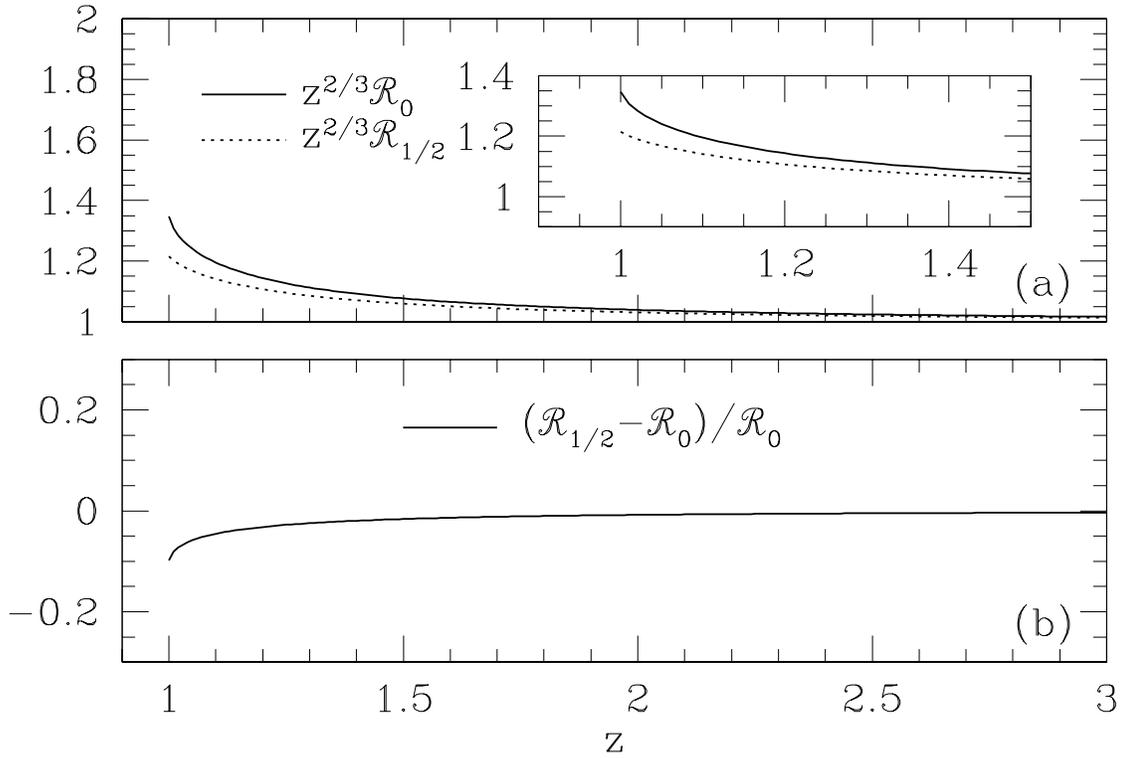}
\caption{ 
Basic functions which describe the photometric properties of
finite sources on the $u_2$-axis (perpendicular to the axis of symmetry of the cusp), as a function of
the distance $z$ in units of the dimensionless source size $\rho_*$. (a) 
The solid line shows $z^{2/3}\cR_0(z)$, whereas the dotted line shows $z^{2/3}\cR_{1/2}(z)$.  
The fractional deviation of a uniform source from the point source
magnification is $z^{2/3}\cR_0 (z) -1$.  (b) The basic function 
describing the photometric behavior of limb-darkened sources, 
$(\cR_{1/2} (z) -\cR_{0} (z))/\cR_0 (z)$.
}
\label{fig:fperp}
\end{figure}

Finally, we can consider the magnification of small sources on the $u_2$-axis.  
Using equation (\ref{eq-RSN}), for a uniform source we find that
$$\cR_{\Sn}^{\rm us} (z) = \cR_0 (z),$$
so
\beq
\label{eq-mlocus-out-u2-us}
\muloutperpus (z)
= \left(\frac{u_{r2}}{\rho_*}\right)^{2/3} \cR_0 (z).
\eeq
For a limb-darkened source, we obtain,
$$\cR_{\Sn}^{\rm ld} (z) = \cR_0 (z) + \Gamma \ [\cR_{1/2} (z) - \cR_0 (z)],$$
and, thus,
\beq
\label{eq-mlocus-out-u2-ld}
\muloutperpld (z)
= \left(\frac{u_{r2}}{\rho_*}\right)^{2/3} 
\left( \cR_0 (z) + \Gamma \ [ \cR_{1/2} (z) - \cR_0 (z) ]\right).
\eeq
Figure \ref{fig:fperp} shows the functions $\cR_0(z)$ and $\cR_{1/2}(z)$, along
with their fractional difference $(\cR_{1/2} (z)-\cR_0 (z))/\cR_0 (z)$.

\section{A Worked Example: Binary Lensing Event MACHO-1997-BUL-28\label{sec:mb9728}}

In this section we numerically calculate the
observable microlensing properties for a binary-lens cusp caustic
crossing.   We do this in order to illustrate the photometric and astrometric 
lensing behavior near a cusp, and to provide an estimate of the magnitudes
of the effects of finite sources and limb darkening for a general
source trajectory.  In the next section (\S\ref{sec:compare}), we
verify and explore the accuracy of the
analytic formulae for the total magnification and centroid derived
in the previous sections by comparing them to the numerical results
obtained here.  We calculate the photometric and astrometric behavior for
the Galactic binary microlensing event {\event}, which was densely
monitored by the PLANET collaboration \citep{albrow1999a}.
The best-fit model for this event has the source crossing an isolated
cusp of a binary lens; the dense sampling near the cusp crossing
resulted in a determination of not only the dimensionless source size
$\rho_*$, but also limb-darkening coefficients in both the $I$- and
$V$-bands.  Combined with an estimate of the angular size $\theta_*$
of the source from its color and apparent magnitude, the measurement
of $\rho_*$ yields the angular Einstein ring radius, 
$\thetae=\theta_*/\rho_*$.  Therefore, the absolute angular scale of the 
astrometric centroid shift is known, and thus, up to an orientation on 
the sky and subject to small errors in the inferred parameters, the 
astrometric behavior can be essentially completely determined from the 
photometric solution.

\subsection{Formalism and Procedures\label{sec:formalism}}

Consider a lens consisting of two point masses located
at positions $\bt_{l,1}$ and $\bt_{l,2}$, with no smoothly
distributed matter or external shear.  In this case,
the dimensionless potential (\Eq{eqn:potential}) is given by,
\beq
\psi (\bt)= m_1 \ln{|\bt-\bt_{l,1}|} + m_2 \ln{|\bt-\bt_{l,2}|},
\label{eqn:potbin}
\eeq
where $m_1$ and $m_2$ are the masses of the two components of the
lens in units of the total mass.  Note that all angles in \eq{eqn:potbin}
are in units of the total mass of the system.  Since $\bma = {\vect{\nabla}} \psi$,
the lens equation (\Eq{eq-le}) becomes,
\begin{equation}
\bu =\bt - 
m_1{{\bt - \bt_{l,1}} \over {|\bt - \bt_{l,1}|^2}}-
m_2{{\bt - \bt_{l,2}} \over {|\bt - \bt_{l,2}|^2}}.
\label{eqn:blenseq}
\end{equation}
Equation (\ref{eqn:blenseq}) is
equivalent to a fifth-order polynomial in $\bt$, thus
yielding a maximum of five images.  All of the image positions for a
given point on the source plane can be found numerically using any
standard root finding algorithm.  Then the individual magnifications
can be found using \eq{eqn:magimagei}, and the 
total magnification and centroid of these images are given by
equations (\ref{def-mtot}) and (\ref{def-colc}).

To calculate the total magnification and centroid for a finite 
size source, it is necessary to integrate the corresponding point-source
quantities over the area of the source (see Eqs.\ \ref{eqn:fsgen}
and \ref{eq-ic-1}).  This is generally difficult or time-consuming
to do directly in the source plane, because of the divergent nature of 
these quantities near the caustic curves; furthermore, it
requires one to solve \eq{eqn:blenseq} for each source position. 
Instead we use the method
of inverse ray shooting, which 
is generally easier, quicker, and more robust. We
sample the image plane uniformly and densely, and use \eq{eqn:blenseq}
to find the source position $\bu(\bt)$ corresponding
to each trial image position $\bt$.  The resulting
source positions are binned, with the magnification
of each bin equal to local ratio of the density of
rays in $\bt$ to the density of rays in $\bu$.  
The result is a map of the magnification $\mu(\bu)$, the
magnification as a function of $\bu$.  Similarly, one can determine
the astrometric deviation by sampling in $\bt$, using \eq{eqn:blenseq}
to determine $\bu(\bt)$, and then summing at each $\bu$ the values of
$\bt(\bu)$.  The astrometric deviation at $\bu$ is then the summed
values of $\bt(\bu)$, weighted by the local magnification.  Thus, one
creates two astrometric maps, for each direction.  
The resolution of the resulting maps is given by
the size of the bins in the source plane.  
Since inverse ray shooting conserves flux, the maps can 
be convolved with any source profile to produce the finite source 
photometric and astrometric behavior for arbitrary source size 
and surface brightness profile.

In this method, the accuracy of a resolution element is limited 
by Poisson fluctuations in the number of rays per bin.  The
fractional accuracy in the magnification $\mu_k$ of resolution element $k$ 
is therefore
\beq
\left({\sigma_{\mu}\over \mu}\right)_k = \frac{1}{\sqrt{\mu_k \sigma \Omega}}
\eeq
where $\sigma$  is the surface of density of rays in the
image plane  and $\Omega$ the area of each resolution
element.   
For a source size encompassing many resolution elements, the total error is 
\beq
\left({{\sigma_\mu}\over \mu}\right)=\left[\dsum_k w_k\left(\frac{\sigma_\mu}{\mu}\right)_k^{-2}\right]^{-1/2},
\eeq
where the sum is over all the resolution elements, and $w_k$ is the 
weight of each resolution element.  For the simulations of
event \event, we require that the resolution of the maps be
considerably smaller than the dimensionless source size $\rho_*$, 
and we require extremely high accuracy in order to explore the
subtle effects of limb-darkening.   For \event, the best-fit model\footnote{
We will be focusing on the first ``LD1'' model of \citet{albrow1999a}, with
parameters given in column two of their Table 1.  This 
corresponds to the best-fit model assuming a one-parameter limb-darkening
law and DoPHOT error bars.  Although their model LD2 provides a marginally better fit, the differences are not important for this study.} 
has $\rho_*=\theta_*/\thetae=0.0287$.  We use a resolution element
of angular size $2 \times 10^{-4}\thetae\simeq 0.007 \ \theta*$ and, thus, 
$\Omega = 4\times 10^{-8}\thetae^2$.  There are approximately $6.5\times 10^4$ resolution
elements per source size.  We sample the image plane with a density
of $\sigma\simeq 10^9 \thetae^{-2}$ and, hence, $(\sigma_\mu/\mu)_k=15\% \mu_k^{-1/2}$.
Therefore, the total fractional error for each source position is always 
$\la 0.1\%$, considerably smaller than any of the effects we are
considering.

\subsection{Global Astrometric Behavior}

\begin{figure}
\plotone{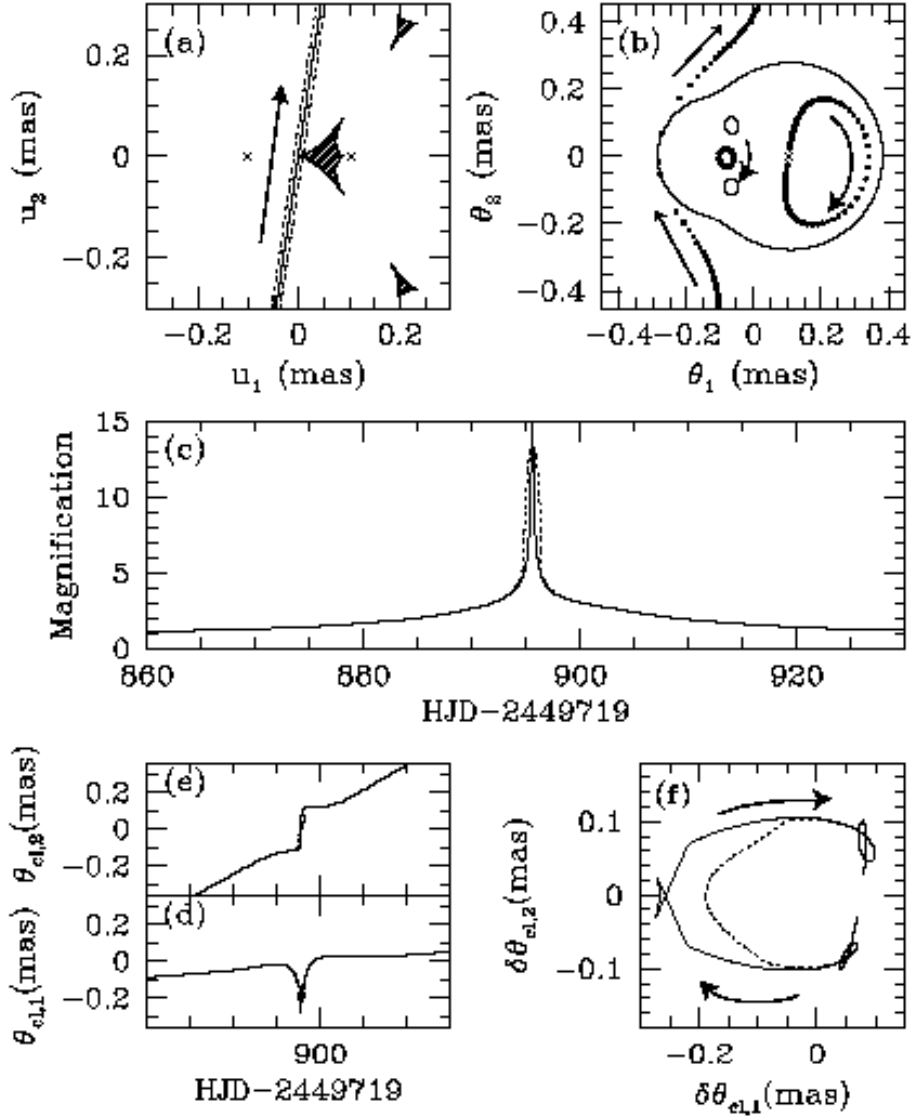}
\caption{ 
The global photometric and astrometric behavior of the best-fit model
for the cusp-crossing Galactic binary-lens microlensing event \event.
(a) The shaded areas indicate the regions interior to the binary-lens
caustic, the solid line shows the trajectory of the center of the source, 
whereas the dotted lines show the width of the source.  The {\sf x}'s 
denote the positions of the two masses of the lens; the heavier mass
is on the right.  (b) The thin, dotted lines show the critical curves. 
The dots show the positions of the images at fixed intervals of 1 day. 
(c) The light curve (magnification as a function of time) for a point source (solid line) and uniform source (dotted line).  (d,e) The two components
of $\btcl$ as a function of time.  The solid line is for a point source, dotted line for a uniform source.  (f) The centroid $\delta \btcl$ of all the images relative to the position of the source.
}
\label{fig:figglo}
\end{figure}

We first analyze the global photometric and astrometric behavior
of event \event, which is summarized in Figure \ref{fig:figglo}. The best-fit model of \citet{albrow1999a} has
the source being lensed by a close-topology binary lens.  
The topology of a binary-lens is specified by the mass ratio, $q=m_2/m_1=0.232$ and
the projected separation in units of the total $\thetae$ of the system, 
$d=|\bt_{l,1}-\bt_{l,2}|=0.687$.  
Furthermore, we shall employ a coordinate
system  such that the masses are located on the $\theta_1$-axis, 
with the
origin equal to the midpoint of the two masses and the heavier
mass being $m_1$, which  is located on the right.  Thus, 
we have $\bt_{l,1}=(d/2,0)$
and $\bt_{l,2}=(-d/2,0)$.  For these parameters $(d,q$), the caustics
form three separate curves, with the largest (primary) caustic
located between the two masses, having four cusps, two of which have
axes of symmetry that are coincident with the $\theta_1$-axis.
Since $m_1\ne m_2$, the caustic is not symmetrical about the $\theta_2$-axis,
and there is a relatively isolated cusp located approximately at the origin. 
See Figure \ref{fig:figglo}(a), where we show the 
caustics and source trajectory for the {\event} model.

For source positions that are large in comparison to $d$, the lens
acts like a single mass located at the center of mass of the binary.
Thus, for such a close binary lens, two of the images behave similarly
to those formed by a single lens.  This can be seen in Figure
\ref{fig:figglo}(b), where we plot the critical curves and image
trajectories for the event.  Similarly, the total magnification
behaves like the magnification of a single lens well away from closest
approach to the origin (midpoint of the binary).  It is only when the
source moves relatively close to the lens that the binary nature of
the lens is noticeable.  See Figure \ref{fig:figglo}(c), where we plot
the magnification as a function of time for 70 days centered on the
maximum magnification of the event.  Inspection of Figure
\ref{fig:figglo}(c) shows that, at the time of the closest approach to
the origin of the lens, the source passes very close ($0.003\thetae$)
to the origin.  It thus traverses the isolated cusp of the primary
caustic, resulting in a spike in the magnification (Fig.\
\ref{fig:figglo}c).  This also results in large excursions in both
components of the image centroid, as can be seen in Figure
\ref{fig:figglo}(c,d,e), where we plot the two components of $\btcl$
as a function of time (panels c and d), and also the trajectory of
centroid shift $\delta \btcl$ (panel e).  Note that $\thetae\simeq 300
{\mu \rm as}$ \citep{albrow1999a}, which sets the scale for $\btcl$ and
$\delta \btcl$.

\subsection{The Cusp Crossing}

Dense photometric sampling of the event allowed \citet{albrow1999a} to
not only determine the global parameters of the lens, but also the
source size $\rho_*$ and limb-darkening coefficients $\Gamma$ in two
different photometric bands, $\Gamma_I=0.67$ ($I$-band) and
$\Gamma_V=0.87$ ($V$-band).  See \eq{eq-ld} for the definition of
$\Gamma$.  We can therefore explore not only the photometric but also
the {\it astrometric} behavior of the cusp crossing, including both
the finite source size and limb darkening.  Furthermore, since
$\thetae$ is approximately known, we can determine the absolute scale of
these astrometric features and assess their detectability by comparing
them with the expected accuracy of upcoming interferometers.

\begin{figure}
\plotone{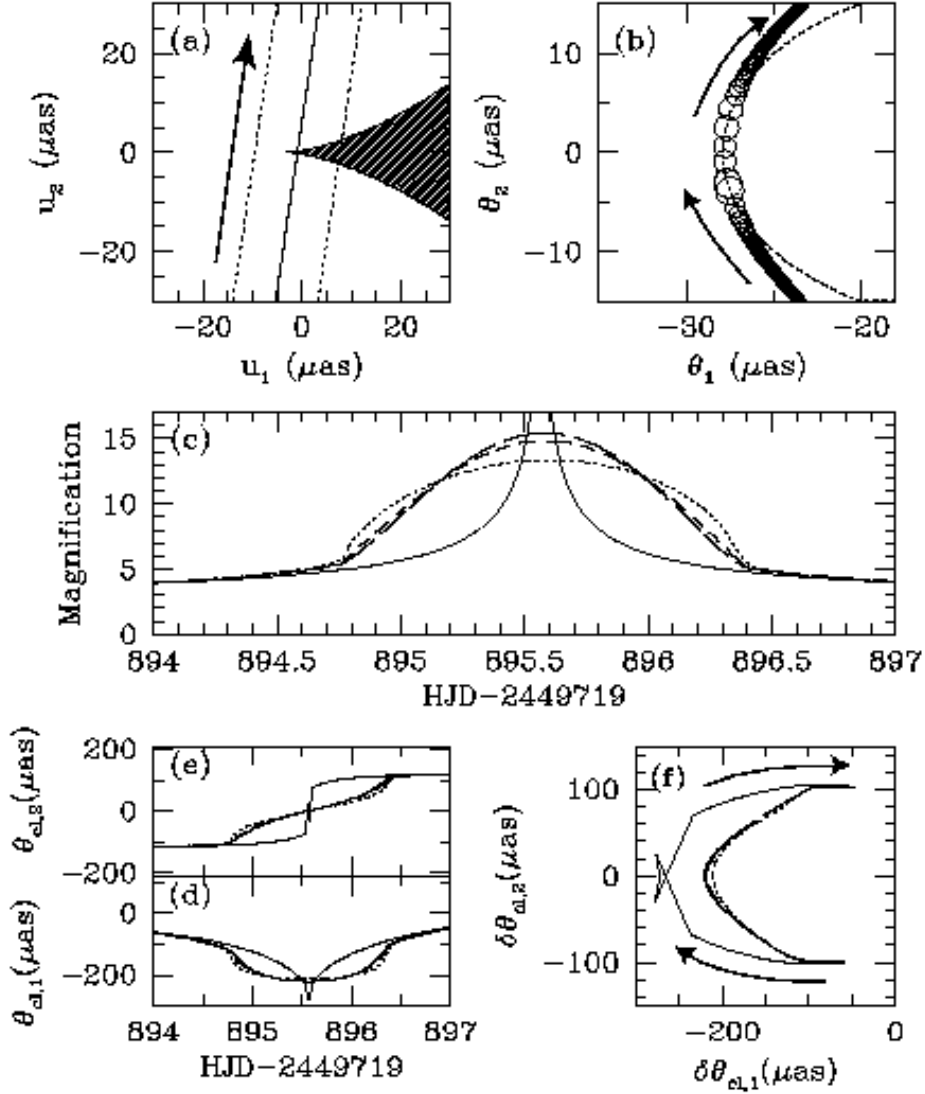}
\caption{ 
The photometric and astrometric behavior of {\event} near the cusp crossing.
(a) The shaded area is shows the region interior to the caustic.  The solid line shows the source trajectory, and the dotted line shows the width of the
source.  (b) The dotted line shows the critical curve, the circles are the positions of the images at fixed intervals of 15 minutes.  The size of each circle is proportional to the logarithm of the magnification of the image.  (c) The magnification
near the cusp crossing as a function of time.  The solid line is for a point
source, dotted line for a uniform source, the short-dashed line for limb-darkened source in the $I$-band, and long-dashed line for the $V$-band.  (d,e)  The two components of the centroid $\btcl$ as a function of time.  Line types are the same as (c).  (f) The path of the centroid shift $\delta\btcl$. 
}
\label{fig:figloc}
\end{figure}

Figure \ref{fig:figloc} summarizes the astrometric and photometric
properties near the cusp crossing.  Figure \ref{fig:figloc}(a) shows
the cusp and source trajectory; the cusp point is located at
$u_{1,c}=-0.009597$, and at closest approach, the center of the lens
is separated by $\sim 0.0066\ \thetae$, or $\sim 20\%$ of a source
radius, from this point.  Figure \ref{fig:figloc}(b) shows the
positions of the images associated with the cusp at intervals of 15
minutes, along with the parabolic critical curve.  The light curve near
the cusp crossing is shown in Figure \ref{fig:figloc}(c), for a point
source, uniform source, and limb-darkened sources in $I$ and $V$.  The
difference between the point source and uniform source is
considerable; whereas the difference between the uniform and
limb-darkened source is smaller (but still significant).  In Figure
\ref{fig:figloc}(d,e), we show the two components of $\btcl$ as a
function of time for the same source profiles, and in Figure
\ref{fig:figloc}(f), we show the trajectory of $\delta\btcl$.  While
the nearly instantaneous jumps are smoothed out by the finite source
size, and also magnitude of the deviations are somewhat suppressed,
$\sim 200\muas$ excursions due to the cusp are still present.  Such
variations should be readily detectable using upcoming
interferometers for sufficiently bright sources.

\begin{figure}
\plotone{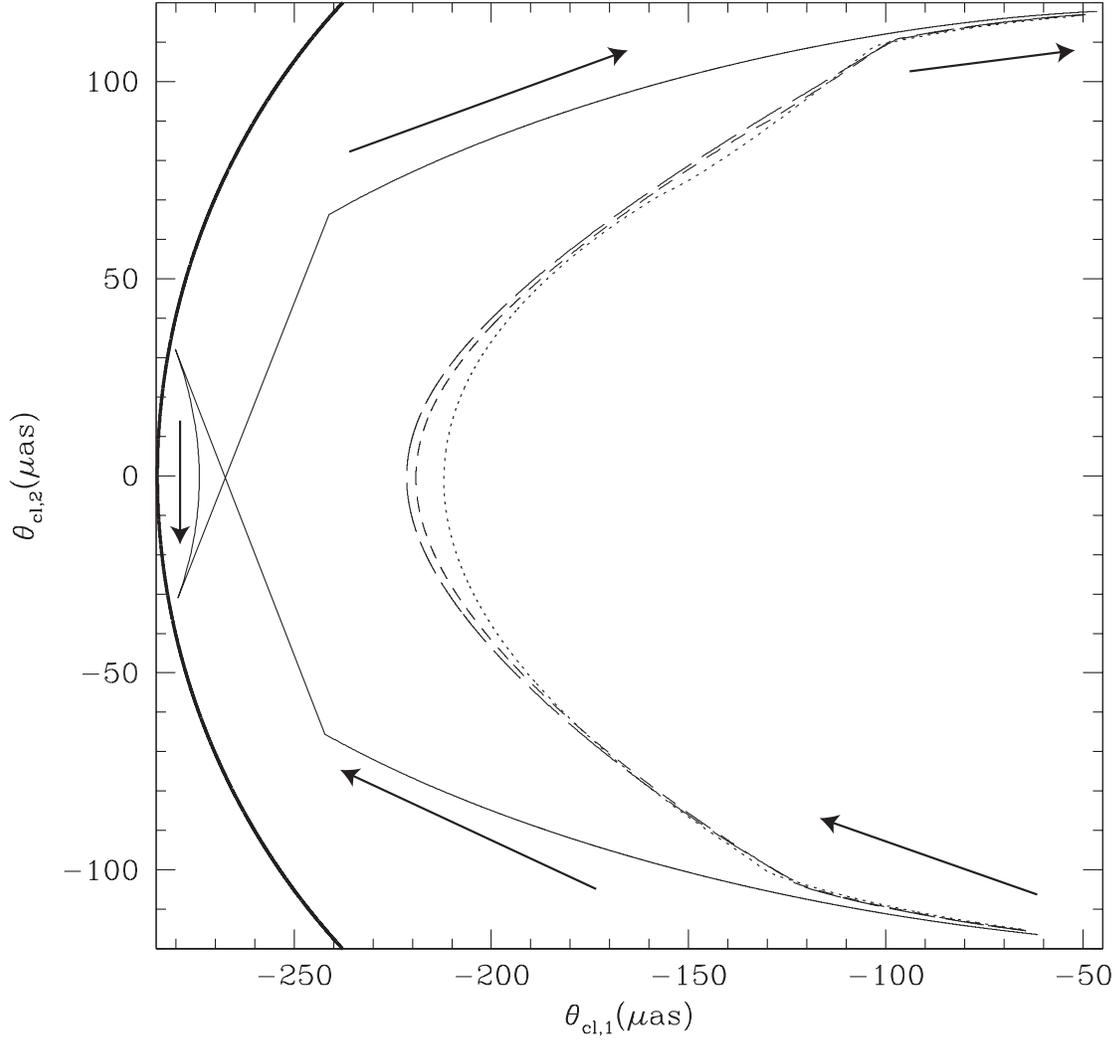}
\caption{ 
Detail of the path of the centroid $\btcl$ of {\event} during a three-day
span centered on the cusp crossing.  The solid line is for a point
source, dotted line for a uniform source, the short-dashed line for limb-darkened source in the $I$-band, and long-dashed line for the $V$-band.
}
\label{fig:figcen}
\end{figure}

We show the path of the centroid $\btcl$ near the cusp in detail in Figure
\ref{fig:figcen}, for a point source, uniform source,
and limb darkened sources with $\Gamma_I=0.67$ and $\Gamma_V=0.87$.
As the source crossed the bottom fold, two images appear at the critical
curve on the opposite side of the axis of symmetry of the cusp from the source, 
resulting in a large, discontinuous jump in $\btcl$.  Interior
to the caustic, there are three images and the 
centroid moves in the opposite direction to the motion of the source, 
until a pair of images disappear when the source is on the top
fold, resulting in another discontinuous jump in $\btcl$.  
Again, these images disappear on the critical curve on the opposite side of the axis as the source.
This motion results in a 
characteristic ``swallowtail'' trajectory that can also be seen in 
Figure \ref{fig:example3}, and is characteristic of 
the centroid of a pointlike source crossing both of the fold arcs
associated with a cusp.  For a sufficiently large source size, as is the case
for \event, this
behavior is completely washed out, resulting in a smooth
trajectory.  Not surprisingly, the maximum astrometric deviation is also
somewhat suppressed.  In this case, the maximum difference between
the astrometric shifts for point source and finite source amounts to 
$\sim 50 \muas$.  

\begin{figure}
\plotone{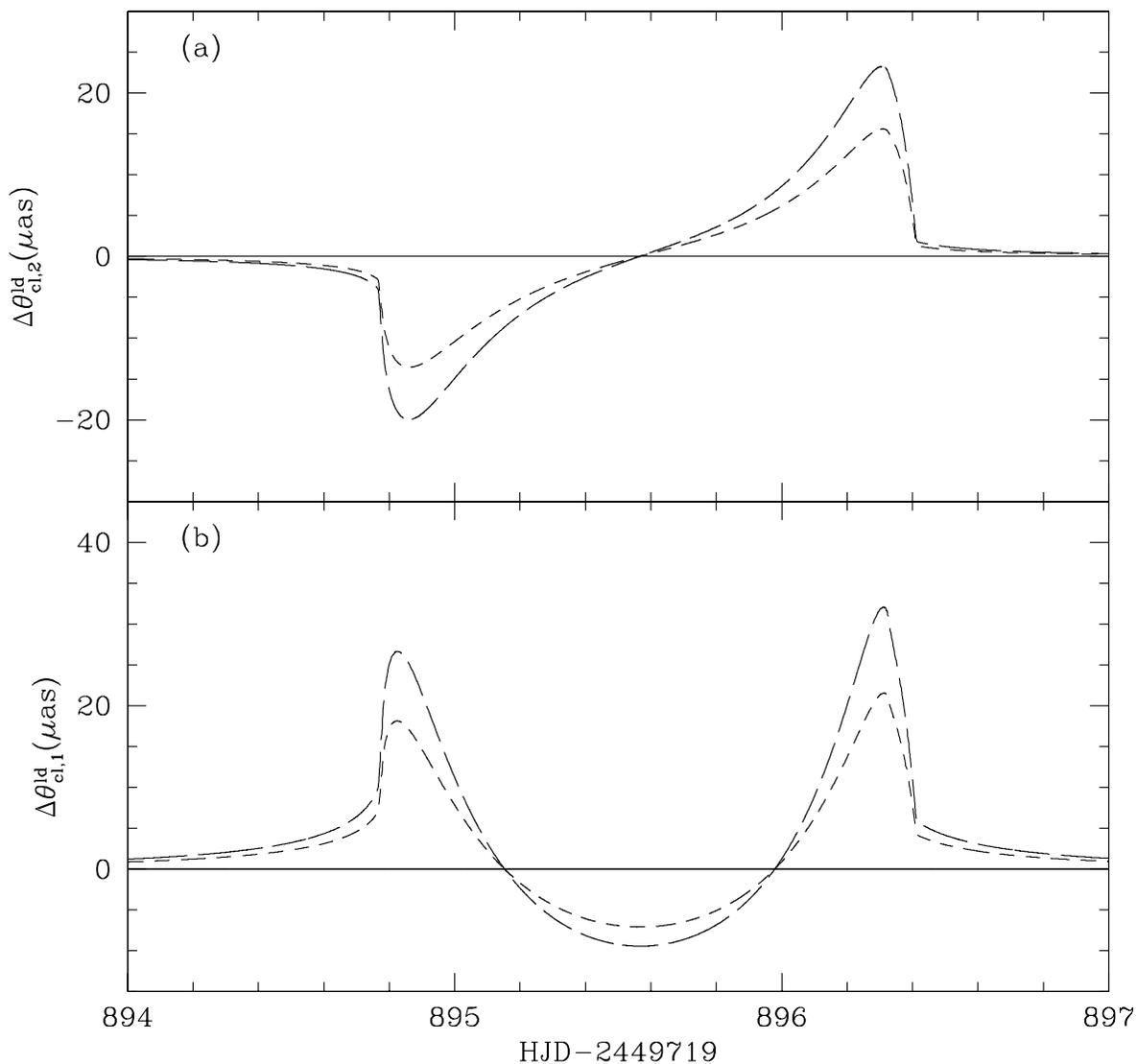}
\caption{ 
(a,b) The two components of the astrometric offset due to limb-darkening $\Delta\bt_{\rm cl}^{\rm ld}=\bt_{\rm cl}^{\rm ld}-\bt_{\rm cl}^{\rm us}$ relative
to a uniform source as a function of time.  The short-dashed line is for the
$I$-band, whereas the long-dashed line is for the $V$-band.  The two 
components are (a) parallel and (b) perpendicular to the axis of the cusp
(i.e., tangent line to the cusp), which is coincident with the binary axis. 
}
\label{fig:figld}
\end{figure}

\begin{figure}
\plotone{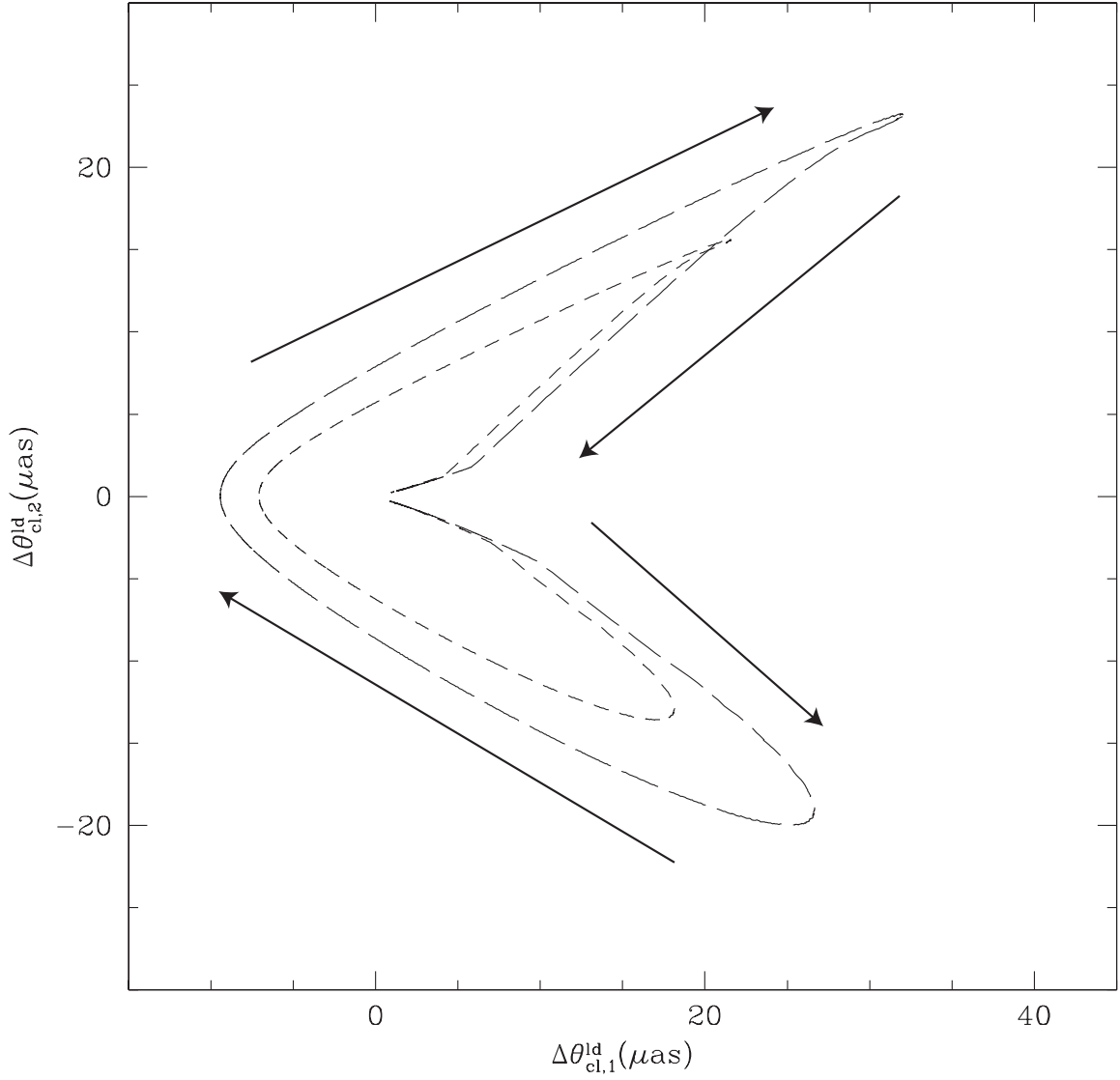}
\caption{The path of the astrometric offset due to limb-darkening $\Delta\bt_{\rm cl}^{\rm ld}=\bt_{\rm cl}^{\rm ld}-\bt_{\rm cl}^{\rm us}$ relative
to a uniform source.
}
\label{fig:figld2}
\end{figure}

In Figure \ref{fig:figld}, we show the two components of $\Delta \bt_{\rm cl}^{\rm ld}\equiv \bt_{\rm cl}^{\rm ld}-\bt_{\rm cl}^{\rm us}$, the astrometric
offset due to limb darkening, as a function of time.  The are two important points to note.  First, the deviation is largest just after the source limb 
first crosses the fold caustic, and just before the source limb leaves 
the caustic entirely.  Second, the two components of $\Delta \bt_{\rm cl}^{\rm ld}$ are not correlated, implying the limb-darkening offset is truly two-dimensional.  This is opposed to the astrometric offset in pure fold caustic
crossings, which is essentially one-dimensional, at least for simple
linear folds (Paper I). This latter point is well-illustrated in Figure 
\ref{fig:figld2}, where we show the trajectory of 
$\Delta \bt_{\rm cl}^{\rm ld}$.   The maximum astrometric shift 
due to limb-darkening for {\event} is $\sim 35\Gamma\muas$.  This
is similar to the maximum astrometric
shift due to limb-darkening for the fold caustic-crossing event
OGLE 1999-BUL-23 in Paper I ($\sim 60\Gamma \muas$), once the factor
of $\sim 2$ difference between the value of $\thetae$ of the two
events is taken into account.   However in 
the case of the cusp, the deviation is $\ga 10\muas$ for a considerable
fraction of the time the source is resolved, whereas in the case of the fold,
the deviation is only $\ga 10\muas$ for the last $\sim 10\%$ of the caustic
crossing.

\section{Comparison with Analytic Expressions\label{sec:compare}}

In \S~\ref{sec:analytic}, we derived analytic expressions for the local
photometric and astrometric behavior of sources near cusp
singularities, using the generic form for the cusp mapping, whereas
in \S\ref{sec:mb9728}, we used numerical techniques to study the
behavior of microlensing observables near the cusp of a binary-lens.
It is important to make contact between the analytic and numerical results,
and, in particular, to explore the validity and applicability of the
analytic results to observable lens systems.

In order to compare the two results, we must specify the coefficients
$a,b$, and $c$ of the generic cusp mapping \eq{eq-le-cusp}, that
correspond to the cusp that was crossed in the binary-lens event
\event.  These coefficients are defined in terms of the local
derivatives of the binary-lens potential at the point on the critical
curve corresponding to the image of the cusp point.  The critical curves of a binary lens can generically be found by solving a quartic complex polynomial \citep{witt1990}.  However, for the binary-lens coordinate system described in the previous section (masses on the $\theta_1$-axis and origin at the midpoint of
the lenses), the equation for critical points on the $\theta_1$ axis
takes the simpler form,
\beq
m_1\left(-\frac{d}{2} - \theta_{c,1}\right)^2+ m_2\left(\frac{d}{2} - \theta_{c,1}\right)^2
-\left(\frac{d}{2} -\theta_{c,1}\right)^2\left(-\frac{d}{2} -\theta_{c,1}\right)^2=0.
\label{eqn:critbin}
\eeq
For $d=0.687$ and $q=m_1/m_2=0.232$, the cusp critical point of interest is
$\bt_{c}=(-0.9488,0)$, which is mapped to  the cusp caustic point
$\bu_{c}=(-0.0096,0)$
using the binary-lens equation (\Eq{eqn:blenseq}).    

The coefficients $a,b,$ and $c$ can be found using the expressions (\Eq{eqn:coeff})
by taking the derivatives of the potential (\Eq{eqn:potbin}).  We find, for 
critical points on the $\theta_1$-axis, that
\beqa
a &=& \frac{m_1}{\left(\theta_{c,1}-d/2\right)^4}+\frac{m_1}{\left(\theta_{c,1}+d/2\right)^4},\\
b &=& \frac{2m_1}{\left(\theta_{c,1}-d/2\right)^3}+\frac{2m_1}{\left(\theta_{c,1}+d/2\right)^3},\\
c &=& 1+ \frac{m_1\left(\theta_{c,1}+d/2\right)^2 + m_2\left(\theta_{c,1}-d/2\right)^2}
{\left(d/2 -\theta_{c,1}\right)^2\left(d/2 +\theta_{c,1}\right)^2}.\label{eqn:cbin}
\eeqa
Inserting \eq{eqn:critbin} into \eq{eqn:cbin} we find that $c=2$ for 
any critical points on the $\theta_{1}$-axis.
For
$\bt_{c}=(-0.9488,0)$, we find $a=1.694$, $b=-2.450$, $c=2$. With
these coefficients, the local behavior of the cusp is completely
specified, and the magnifications and positions of the images
associated with the cusp can be determined using the expressions
derived in \S~\ref{sec:local}.  However, in order to compare with the
photometric and astrometric behavior of \event, not only the local
behavior of the cusp images must be specified, but also the behavior
of the images not associated with the cusp.  As in
\S~\ref{sec:global}, we assume that the magnification and centroid of
the two images unrelated to the cusp can be well-represented by a
first-order Taylor expansion about the time $\tcc$ when the source
crosses the axis of the cusp (i.e., tangent line to the
cusp).  We find the slope and intercept of these
expansions that minimizes the differences between the analytic and
numerical calculations.  Note that these parameters (the three slopes
and three intercepts) are the only free parameters, the photometric
and astrometric behavior of the images associated with the cusp are
completely determined once the coefficients $a,b$, and $c$ are
specified.

\begin{figure}
\plotone{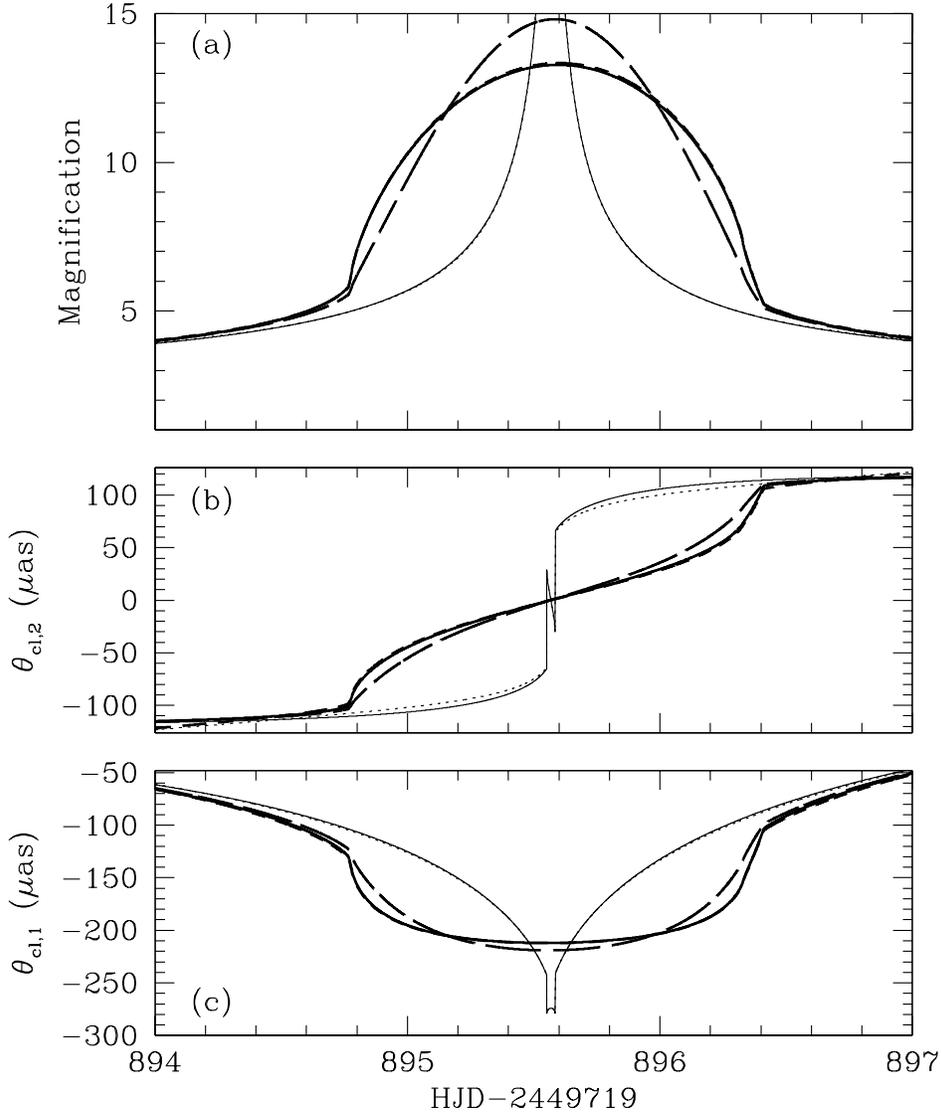}
\caption{ 
A comparison between the photometric and astrometric properties of the
event {\event} near the cusp crossing computed using the full binary-lens formalism, and using the generic cusp mapping.  The coefficients of the cusp mapping were determined from the local derivatives of the potential at the position of the image of the binary-lens cusp.  See text.  (a)  The light solid line shows the point-source magnification as a function of time calculated using the binary-lens formalism, the
light dotted line (barely visible) is using the analytic expressions derived here.  The heavy solid line shows the uniform-source magnification calculated
using the full binary-lens equation, the short-dashed line is using the generic cusp mapping.  For comparison, the long-dashed line is the magnification for
the $I$-band.  (b,c) The two components of the centroid $\btcl$.  Line types are the same as in panel (a).   
}
\label{fig:compare}
\end{figure}

In Figure \ref{fig:compare}, we show the point-source photometric and
astrometric behavior near the cusp crossing of \event, as determined
using both the full binary-lens formalism and using the analytic
expressions.  Also shown in Figure \ref{fig:compare} 
are the corresponding predictions for a finite source.  To calculate the
finite source magnification and centroid for the generic cusp, we adopt the
inverse ray-shooting procedure described in \S~\ref{sec:formalism}, using
the cusp mapping (\Eq{eq-le-cusp}) to predict
the source positions.  In both the point-source
and finite source case, the agreement is quite remarkable.  We find that the
difference in magnification is $\la 5\%$, and the difference in the
centroid is $\la 5 {\muas}$.  This is generally smaller than any of the
physical effects we have considered, including limb-darkening.

\section{Some Applications\label{sec:apply}}

Generally speaking, the results derived in \S\ref{sec:analytic} on the
photometric and astrometric microlensing properties near cusps are
useful for two basic reasons.  First, they are (for the most part)
analytic, making computations enormously simpler and considerably less
time-consuming, and making results easier to interpret.  This allows
for rapid and efficient exploration of parameter space; a significant
advantage over numerical methods.  Second, the results are generic;
they are applicable to any cusp produced by a gravitational lens.
Similarly, they are local, being tied to the global lens in
consideration only via the coefficients $a,b,c$ of the generic cusp
mapping (\Eq{eq-le-cusp}).  This means that the observables due to the
cusp crossing can be analyzed in a parametric manner (with the
coefficients $a,b,c$ being the parameters), without regard to the
global (and presumably, {\it non-analytic}) properties of the lens in
question.  In this section, we discuss several possible applications
of our results that take advantage of these properties.

\begin{figure}
\plotone{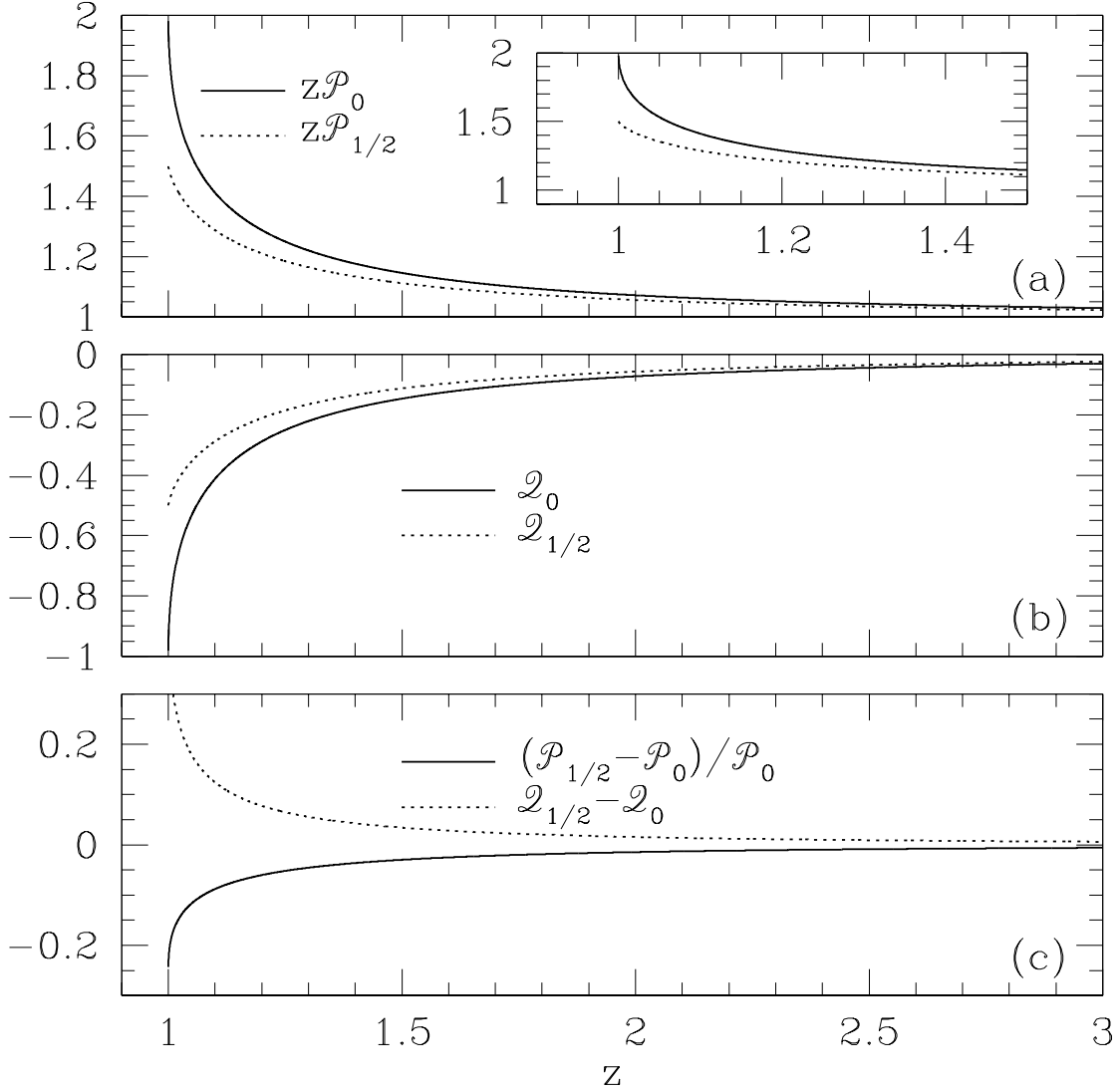}
\caption{ 
Basic functions which describe the photometric and astrometric properties of
finite sources on the $u_1$-axis (the axis of symmetry), as a function of
the distance $z$ in units of the dimensionless source size $\rho_*$. (a) The basic functions for the photometric behavior.  The solid line shows $z\cP_0(z)$, whereas the dotted line shows $z\cP_{1/2}(z)$.  
The fractional deviation of a uniform source from the point source
magnification is $z\cP_0 (z) -1$.  (b) The basic functions for the astrometric
behavior. The solid line shows $\cQ_0(z)$; the dotted line shows $\cQ_{1/2}(z)$.  
The astrometric deviation from a point source is proportional 
to $\cQ_{0}$ for a uniform source.  (c)  The basic functions describing the 
photometric ($(\cP_{1/2} (z) -\cP_{0} (z) )/\cP_0 (z)$; solid) 
and astrometric ($\cQ_{1/2} (z) -\cQ_{0} (z)$; dotted) behavior 
of limb-darkened sources.
}
\label{fig:fpar}
\end{figure}

One issue often encountered in microlensing studies is the question of
when the point source approximation is valid.  This is important,
because it is typically easier to calculate the observable
properties under the assumption of a point-source than when a finite
sources is considered.  For a given source location, the accuracy of
the point source approximation depends primarily on the distance of
the source from the nearest caustic, relative to the size of the
source, and secondarily on the surface brightness profile of the
source.  For a source near a simple linear fold, there exist
semi-analytic expressions for the finite source magnification,
relative to the point-source magnification, for uniform and
limb-darkened sources (see, e.g.\ Paper I).  For a source
interior to the fold caustic (where there are an additional two images
due to the fold), the difference in the uniform source magnification
from the point source magnification is $\la 5\%$ for separations $z\ga
1.5$ , where $z$ is the distance of the center of the source from the
fold in units of the source radius, and the point-source magnification
is proportional to $z^{-1/2}$.  The effect of limb-darkening is $\la
1\%\Gamma$ in the same range, where $\Gamma$ is the limb-darkening
parameter, which has typical values in the range $0.5-0.7$ for optical
wavelengths.  Using the results derived in \S\ref{sec-us-limb}, we can
also determine the effect of finite sources on the magnification near
a cusp.  For sources outside the cusp and on the axis of the cusp, the point
source magnification is $\mulout = u_{r1}/(\rho_* z)$
(\Eq{eq-mag-out-u1-axis}), whereas the uniform source magnification is
$\mulusout = (u_{r1}/\rho_*)\cP_0(z)$ (\Eq{eq-mlocus}).  Thus, the
fractional deviation from a point source is $z\cP_0(z)-1$, independent
of $\rho_*$ and $u_{r1}$ (at least for small sources).  Figure
\ref{fig:fpar}(a) shows $z\cP_0(z)$ as a function of $z$.  We find
that the uniform source magnification deviates from the point source
magnification by $\la 5\%$ for $z \ga 2.5$.  Similarly, one can show
that the effects of limb-darkening are $\la 1\%\Gamma$ for $z \ga
2.5$;  see \eq{eq-mag-fsloc} and Figure \ref{fig:fpar}(c).  For a given
distance to the cusp, the effects of finite sources are less severe
for sources perpendicular to the axis of the cusp than parallel to the axis
(see equations [\ref{eq-mlocus-out-u2-us}] and
[\ref{eq-mlocus-out-u2-ld}]).  Thus, one can use the results for
sources along the axis of the cusp to determine the applicability of the
point-source approximation for the magnification given a certain
photometric accuracy and arbitrary source trajectory relative to the
axis of the cusp.  It is interesting to note that, at a given distance from the
caustic, the effects of finite uniform and limb-darkened sources are
larger for cusps than for folds.

The remarkable agreement of the analytic predictions for the
properties of \event~ with the exact numerical calculation suggests
another application of our results.  One primary difficulty with
modeling Galactic caustic-crossing binary lenses is the fact that the
magnification is generally non-analytic.  Furthermore, in order to
calculate the finite source magnification, this non-analytic
point-source magnification must be integrated over the source.  This
can be quite time consuming, thus making it difficult to explore quickly and
efficiently the viable region of parameter space.
Furthermore, the observables are generally not directly related to the
canonical parameters.  This means that small changes in the parameters
generally lead to large changes in the observables.  Therefore, the
parameters must be quite densely sampled, further exacerbating the
problems with the calculations.  Thus, finding all viable model fits to
well-sampled Galactic binary lensing events has proved quite
difficult.  These problems have been discussed in the context of
fold-caustic crossing binary lens events in \citet{albrow1999b}, who
proposed an elegant and practical solution.  Taking advantage of the
fact that the behavior near a fold caustic is universal, and using
analytic expressions for the point-source and finite-source
magnification, \citet{albrow1999b} devised a procedure wherein the
behavior near the fold crossing is isolated and fit separately from
the remainder of the light curve.  Again, because the behavior near a
fold can be calculated semi-analytically, fitting this subset of the
data is considerably easier and quicker than fitting the entire
data set.  This fit can then be used to constrain the viable regions of
the parameter space of the global binary-lens fit.  We propose that a
similar procedure might be used to fit cusp-crossing binary-lens
events.  Consider an observed binary-lens event with one well-sampled
cusp crossing (e.g.\ \event).  Extracting the portion of the
light curve near the cusp crossing, a fit to the generic cusp forms
presented in \S\ref{sec:analytic} can be performed, allowing for
images not associated with the cusp.  This fit can then be used to
constrain viable combinations of the parameters $a,b,c$.  One can then
search for cusps in the space of binary-lens models that satisfy
these constraints.  Finally, one can search for viable fits to the
entire data set in this restricted subset of parameter space.  Although
there are undoubtedly nuances in the implementation, it seems likely
that this (admittedly schematic) procedure should provide a relatively
robust and efficient method for fitting many cusp-crossing binary-lens
events.

We discuss one final application of our results that may prove useful
in the context of quasar microlensing.  For poorly-sampled microlensed
light curves, there exists a rough degeneracy between the effects of
the size of the source and the typical angular Einstein ring radius
$\left<\thetae\right>$ (see, e.g.\ \citealt{wwt2000b}).  As discussed by \citet{lw1998}, this
degeneracy can be broken with astrometric observations, since the
scale of the motion of the centroid of all the microimages is set by
$\left<\thetae\right>$.  Due to the highly stochastic nature of the
centroid motion for microlensing at high optical depth, in order to
use an observed data set to determine $\left<\thetae\right>$, one would
have to calculate many realizations of the distribution of centroid
shifts, and compare these statistically to the observed distribution.
This procedure is likely to be quite time-consuming.  It may be
possible to use the results presented here to devise a method to
obtain a cruder, but considerably less time consuming, estimate for
$\left<\thetae\right>$ given an observed data set.  This method makes
use of the analytic results we obtained for fold caustic
crossings. The centroid shift can be quite large when the source
crosses a fold caustic, when two additional images appear in a
position generally unrelated to the centroid of the other images.  In
Paper I, we argued that finite sources only provide a small
perturbation to the centroid jump when the source crosses a fold
caustic, at least for sources with angular size significantly smaller
than $\left<\thetae\right>$.  For the case of a pure fold caustic
crossing, it is not possible to provide a prediction for the magnitude
of the centroid jump from a purely local analysis (Paper I).  However,
in the case of the source crossing one of the two fold arcs abutting a
cusp (see Fig.\ref{fig:example2}), the fact that only two images are
associated with the fold, and one image varies continuously as the
source crosses the caustic, implies that one can make a definite
prediction for the magnitude $|\bphicl^{\rm jump}|$ of the centroid
jump for a given source trajectory for a generic cusp.  In
\S\ref{subsec-jump}, we found that, for sources crossing a fold arc at
a horizontal distance from the cusp of $|u^0_1|$, the centroid jump is
given by, 
\beq 
\left| \bphicl^{\rm jump} \right|
=
\left| \btcl^{\rm jump} \right| \thetae \ 
\simeq \   \sqrt{\frac{u^0_1}{ u_{\rm jump} }} \  \thetae, 
\eeq 
where $u_{\rm
jump}\equiv -(2ac-b^2)/(6b)$.  This expression is valid for $|u^0_1|$
small, and assuming that the total magnification of the images not
associated with the cusp is small.  Given a certain global lens
geometry (i.e.,\ shear and convergence), one can
obtain an estimate of the distribution of the quantity $u_{\rm jump}$
by using known techniques to locate the cusps, and taking the local
derivatives of $\psi$ at the corresponding image points to determine
the coefficients $a,b,c$.  Thus, the distribution of $u_{\rm jump}$ can
be determined.  Similarly, the distribution of $u^0_1$ can be estimated
once the locations of the cusps are known.  Therefore, one can predict
the distribution of $(u^0_1/u_{\rm jump})^{1/2}$ for a given lens
system. Comparing this to an observed distribution of $|\bphicl^{\rm
jump}|$, one can constrain $\left<\thetae\right>$.  We note that there
are some limitations to this method.  It is generally only appropriate
for sources that are small with respect to $\left<\thetae\right>$.  Also,
it assumes that the observed centroid shifts are dominated by caustic
crossings, and that these are well-approximated by the folds abutting
cusps.

\section{Summary and Conclusion\label{sec:summary}}

We have presented a comprehensive, detailed, and quantitative study of
gravitational lensing near cusp catastrophes, concentrating on the
specific regime of microlensing (when the individual images are
unresolved).  We started from a generic polynomial form for the lens mapping
near a cusp that relates the image positions to the source position.
This mapping is valid to third order in the image position.  The
quantitative properties of this mapping are determined by the
polynomial coefficients, which can be related to local derivatives of
the projected potential of the lens.  Near a cusp, the critical curve
is a parabola, which maps to the cusped caustic.  We find an simple
expression for the vertical component of the image position
$\theta_2$, which is a cubic of the form $\theta_2^3 + \ttp \theta_2 +
\ttq=0$, where $\ttp$ and $\ttq$ are functions of the source position
$\bu$.

The solutions of the cubic in $\theta_2$ are characterized by the
discriminant $D(\bu)=(\ttp/3)^3 + (\ttp/2)^2$.  For source positions
$\bu$ outside the caustic, we have $D(\bu)>0$ and, thus, there is locally one
image.  We determined the magnification and location of this image,
and showed, in particular, that along the axis of the cusp
(tangent line to the cusp), the magnification
scales as $u^{-1}$, where $u$ is the distance from the cusp
point. Perpendicular to the axis, the magnification scales as
$u^{-2/3}$.  We also determined the image positions and magnifications
for sources on the caustic, where $D(\bu)=0$.  On the caustic, there
are two images.  One image is infinitely magnified and results
from the merger/creation of a pair of images,
whereas the second image has finite magnification, and can be
smoothly joined to the single lensed image of a source just outside
the caustic.  For sources inside the caustic ($D(\bu)<0$), we
find that there are three images.  One image has positive parity and
diverges as the source approaches the top fold caustic abutting the cusp, but
can be continuously joined to the non-divergent image when the source is on the bottom
fold.  Similarly, the other positive parity image diverges as the source 
approaches the bottom fold, but can be continuously joined to the
non-divergent image when the source is on the top fold.  The third
image has negative parity, and diverges as the source approaches
either fold caustic.  All three images diverge as one approaches the
cusp point.  

For sources on the axis of the cusp, but interior to the
caustic, the total magnification of all the images diverges as
$u^{-1}$.  We also derived analytic expressions for the centroid of all
three images created when the source is interior to the caustic.  We
generalized our results beyond the local behavior near the cusp by
deriving expressions for the total magnification and centroid
including the images not associated with the cusp.  We further
considered rectilinear source trajectories, and parameterized this
trajectory in order to calculated the dependence of the photometric
and astrometric behavior on time.  In particular, due to the presence
of the infinitely magnified images when the source is on the fold
caustics, but finite magnification outside the caustic, we find that
the centroid exhibits a finite, instantaneous jump whenever the source
crosses one of the two folds abutting the cusp.  We present a formula
for the magnitude of the jump that depends only on the local
coefficients of the cusp mapping, and the location of the caustic
crossing.  We note that this magnitude decreases monotonically as
a function of the horizontal distance between the cusp and where
the source crosses the cusped caustic curve,
so that a source which crosses the cusp point exactly exhibits no jump
discontinuity.

Beginning with the appropriate modifications to the formulae for the
total magnification and centroid for finite sources with arbitrary
surface brightness profiles, and combining these with the analytic
results we obtained for the magnification and centroid for point
sources near a cusp, we derived semi-analytic expressions for the
uniform and limb-darkened finite source magnification for small
sources on and perpendicular to the axis of the cusp.  We also
derived expressions for the centroid of a small source on the
axis for uniform and limb-darkened sources.

In order to illustrate the photometric and astrometric lensing
behavior near cusps, and to provide order-of-magnitude estimates for
the effect of finite sources and limb-darkening on these properties,
we numerically calculated the total magnification and centroid shift
for the observed cusp-crossing Galactic binary microlensing event
\event.  We find that the cusp crossing results in large, ${\cal
O}(\thetae)$ centroid shifts, which should be easily detectable with
upcoming interferometers.  We find that limb-darkening induces a
deviation in the centroid of $\sim 35\Gamma \muas$.

We compared our numerical calculations with the analytic expectation,
and found excellent agreement.  Adjusting only the magnification and
centroid of the images unrelated to the cusp, and adopting the
coefficients appropriate to the cusp of \event, we find that our
analytic formulae predict the magnification to $\la 5\%$ and the
centroid to $5\muas$, for positions within $\sim 2$ source radii of
the cusp.

Finally, we suggested several applications of our results to both
Galactic and cosmological microlensing applications.  We suggest that
one can use our analytic expressions to determine the applicability of
the point-source approximation for sources near a cusp.  In
particular, we note that the finite source magnification deviates from
the point-source magnification by $\la 5\%$ for sources separated by
$\ga 2.5$ source radii.  We also discussed how the local and generic
behavior of the cusp can be used to simplify the fitting procedure for
cusp-crossing events.  Lastly, we outlined a method by which the
typical angular Einstein ring radius of the perturbing microlenses of a
macrolensed quasar might be estimated using measurements of the jump
in the centroid that occurs when the source crosses a fold, making use
of the analytic expression for the magnitude of the jump derived here.

Despite their apparent diversity, the mathematical underpinning of all
gravitational lenses is identical.  In particular, all lenses exhibit
only two types of stable singularities: folds and cusps.  In Paper I,
we studied gravitational microlensing near fold caustics; here we have
focussed on microlensing near cusp caustics.  A generic form of the
mapping from source to image plane near these each of these types of
caustics can be found, and used to derive mostly analytic expressions
for the photometric and astrometric behavior near folds and cusps.
These expressions can be used to predict the behavior near all stable
caustics of gravitational lenses, and applied to a diverse set of
microlensing phenomena, including Galactic binary lenses and
cosmological microlensing.

\acknowledgements
We are especially thankful to Meredith Houlton for meticulously
checking our calculations.
B.S.G. was supported in part by NASA through a Hubble Fellowship grant
from the Space Telescope Science Institute, which is operated by the
Association of Universities for Research in Astronomy, Inc., under
NASA contract NAS5-26555.  A.P. was supported in part by an Alfred
P. Sloan Research Fellowship and NSF Career grant DMS-98-96274.

\clearpage

\epsscale{0.9}

\epsscale{0.9}

\epsscale{0.9}

\epsscale{0.8}

\epsscale{1.0}

\epsscale{0.8}

\epsscale{1.0}

\begin{thebibliography}{}


\bibitem[Agol \& Krolik(1999)]{ak1999}
Agol, E., \& Krolik, J.\ 1999, ApJ, 524, 49

\bibitem[Albrow et~al.(1999a)]{albrow1999a} % mb28
Albrow, M., et~al.\ 1999a, ApJ, 522, 1011

\bibitem[Albrow et~al.(1999b)]{albrow1999b} % complete
Albrow, M., et~al.\ 1999b, ApJ, 522, 1022

\bibitem[Albrow et~al.(2000)]{albrow2000} %9741
Albrow, M.~D.~et al.\ 2000, \apj, 534, 894

\bibitem[Alcock et al.(2000)]{alcock2000} Alcock, C., et al.\ 2000,
ApJ, 541, 270 %binaries

\bibitem[Fluke \& Webster(1999)]{fw1999} 
Fluke, C.~J.~\& Webster, R.~L.\ 1999, \mnras, 302, 68

\bibitem[Gaudi \& Gould(1999)]{gandg1999} Gaudi, B.S., \& Gould, A.\
1999, ApJ, 513, 619

\bibitem[Gaudi, Graff, \& Han(2002)]{ggh2002}
Gaudi, B.S., Graff, D.S., Han, C.\ 2002, in preparation

\bibitem[Gaudi \& Petters(2002)]{paper1}
Gaudi, B.S., \& Petters, A.O.\ 2002, ApJ, 574, 000 (Paper I)

\bibitem[Gould(2001)]{gould2001} % stellar atmospheres
Gould, A.\ 2001, PASP, 113, 903

\bibitem[Graff \& Gould(2002)]{gandg2002} %binary microlens mass meas.
Graff, D., \& Gould, A.\ 2001, ApJ, submitted 

\bibitem[Lewis \& Ibata(1998)]{lw1998} 
Lewis, G.~F.~\& Ibata, R.~A.\ 1998, \apj, 501, 478

\bibitem[Mao(1992)]{mao92}
Mao, S.\ 1992, ApJ, 389, 63

\bibitem[Mao \& Paczy\'nski(1991)]{mandp1991}
Mao, S., \& Paczy\'nski, B.\ 1991, ApJ, 374, 37

\bibitem[Petters(1992)]{ptt92} Petters, A. O. 1992,
            J. Math. Phys., 33, 1915.  


\bibitem[Petters(1995)]{ptt95} Petters, A. O. 1995,
            J. Math. Phys., 36, 4276.


\bibitem[Petters et al.(2001)]{plw01} Petters, A. O.,  Levine, H.,
     \& Wambsganss, J. 2001,  Singularity Theory and Gravitational
     Lensing (Boston: Birkh\"auser). 

\bibitem[Schneider et~al.(1992)]{sef92} Schneider, P., Ehlers, J., \& Falco, E. E.,
     1992,  Gravitational Lenses (Berlin: Springer).

\bibitem[Schneider \& Weiss(1992)]{sw92} 
Schneider, P., \& Weiss A., 1992, \aap, 260,1

\bibitem[Wambsganss, Paczynski, \& Schneider(1990)]{wps1990}
Wambsganss, J., Paczynski, B., \& Schneider, P.\ 1990, ApJ, 358, L33

\bibitem[Witt(1990)]{witt1990}
Witt, H.\ 1990, A\&A, 236, 311

\bibitem[Wyithe, Webster, \& Turner(1999)]{wwt1999} %transverse velocity
Wyithe, J.~S.~B., Webster, R.~L., \& Turner, E.~L.\ 1999, \mnras, 309, 261

\bibitem[Wyithe, Webster, \& Turner(2000)]{wwt2000a} %Q2237 mass function
Wyithe, J.~S.~B., Webster, R.~L., \& Turner, E.~L.\ 2000a, \mnras, 315, 51

\bibitem[Wyithe, Webster, \& Turner(2000)]{wwt2000b} %small Q2237 source
Wyithe, J.~S.~B., Webster, R.~L., \& Turner, E.~L.\ 2000b, \mnras, 318, 762

\bibitem[Zakharov(1995)]{zak95} 
Zakharov, A.\ 1995, A\&A, 293, 1

\bibitem[Zakharov(1999)]{zak99} Zakharov, A.,  in
     Proc. of the Eighth Marcel Grossmann Meeting on General Relativity,
     eds. T. Piran and R. Ruffini
     (Singapore: World Scientific). 

\bibitem[Zwillinger(1996)]{zwi96} Zwillinger, D. 1996,
     CRC Standard Mathematical Tables and Formulae, 30th Edition
   (Boca Raton: CRC Press).


\end{thebibliography}
\end{document}